# A Parametric Framework for Reversible $\pi$-Calculi[*]

Doriana Medić
IMT School for Advanced Studies Lucca, Italy
doriana.medic@imtlucca.it

Claudio Antares Mezzina
IMT School for Advanced Studies Lucca, Italy
claudio.mezzina@imtlucca.it

Iain Phillips
Imperial College London, UK
i.phillips@imperial.ac.uk

Nobuko Yoshida
Imperial College London, UK
n.yoshida@imperial.ac.uk

This paper presents a study of causality in a reversible, concurrent setting. There exist various notions of causality in $\pi$-calculus, which differ in the treatment of parallel extrusions of the same name. In this paper we present a uniform framework for reversible $\pi$-calculi that is *parametric* with respect to a data structure that stores information about an extrusion of a name. Different data structures yield different approaches to the parallel extrusion problem. We map three well-known causal semantics into our framework. We show that the (parametric) reversibility induced by our framework is causally-consistent and prove a causal correspondence between an appropriate instance of the framework and Boreale and Sangiorgi's causal semantics.

## 1 Introduction

Starting from the 1970s [5] reversible computing has attracted interest in different fields, from thermodynamical physics [3], to systems biology [14, 26], system debugging [29, 16] and quantum computing [17]. Of particular interest is its application to the study of programming abstractions for reliable systems: most fault-tolerant schemes exploiting system recovery techniques [2] rely on some form of *undo*. Examples of how reversibility can be used to model transactions exist in CCS [13] and higher-order $\pi$-calculus [19].

A reversible system is able to execute both in the forward (normal) direction and in the backward one. In a sequential setting, there is just one order of reversing a computation: one has just to undo the computation by starting from the last action. In a concurrent system there is no clear notion of last action. A good approximation of what is the last action in a concurrent system is given by *causally-consistent* reversibility, introduced by Danos and Krivine for reversible CCS [12]. Causally-consistent reversibility relates causality and reversibility of a concurrent system in the following way: an action can be reversed, and hence considered as a last one, provided all its consequences have been reversed.

In CCS [24], there exists just one notion of causality: so-called *structural* causality, which is induced by the prefixing '.' operator and by synchronisations. As a consequence, there is only one way of reversing a CCS trace, and from an abstract point of view there exists only one reversible CCS. Evidence for this has been given in [22], where an equivalence is shown between the two methods for reversing CCS (namely RCCS [12] and CCSK [27]).

When moving to more expressive calculi with name creation and value passing like the $\pi$-calculus, matters are more complex. As in CCS, structural causality in the $\pi$-calculus is determined by the nesting of the prefixes; for example, in process $\overline{b}a.\overline{c}e$ the output on channel $c$ structurally depends on the output on $b$. Extruding (or opening) a name generates an *object* dependency; for example, in process $va\,(\overline{b}a\mid a(z))$ the input action on *a* depends on the output on *b*. In the case of parallel extrusions of the same name, for example $va\,(\overline{b}a \mid \overline{c}a \mid a(z))$, there exist different interpretations of which extrusion will cause the action $a(z)$. In what follows, we consider three approaches.

---

[*]This work was partially supported by COST Action IC1405 "Reversible computation – extending horizons of computing", EPSRC EP/K034413/1, EP/K011715/1, EP/L00058X/1, EP/N027833/1 and EP/N028201/1.

The classical and the most used approach to causality in the π-calculus is the one where the order of extrusions matters and the first one of them is the cause of the action $a(z)$. Some of the causal semantics representing this idea are [15, 6, 8] and all of them are defined for standard (forward-only) π-calculus. In [15] the authors claim that, after abstracting away from the technique used to record causal dependences, the final order between the actions in their semantics coincides with the ones introduced in [6, 8]. Hence we group these semantics together as a single approach to causality.

Secondly, in [9], action $a(z)$ in the example above depends on one of the extruders, but there is no need to keep track of which one exactly. This causal semantics is defined for the forward-only π-calculus.

Finally, the first compositional causal semantics for the reversible π-calculus is introduced in [10]. In the above example, parallel extrusions are concurrent and the action $a(z)$ will record dependence on one of them (exactly which one is decided by the context). This causal semantics enjoys certain correctness properties which are not satisfied by other semantics.

Here we present a framework for reversible π-calculus that is parametric with respect to the data structure that stores information about an extrusion of a name. Different data structures will lead to different approaches to the parallel extrusion problem, including the three described above. Our framework allows us to add reversibility to semantics where it was not previously defined. By varying the parameters, different orderings of the causally-consistent backward steps are allowed. Our intention is to develop a causal behavioural theory for the framework, in order to better understand different interpretations of reversibility in the π-calculus, and to use this understanding for causal analysis of concurrent programs.

A preliminary discussion about the framework appeared in [23], where some initial ideas were given. Moreover in [23] it was argued the necessity to modify the semantics of [6] in order to add information about silent actions. In this work we fully develop the idea behind the framework and detach from modifying the semantics of [6].

**Contributions.** We present a framework for reversible π-calculus which is parametric in the bookkeeping data structure used to keep track of object dependency. As reversing technique, we will extend the one introduced by CCSK [27], which is limited to calculi defined with GSOS [1] inference rules (e.g., CCS, CSP), to work with more expressive calculi featuring name passing and binders. This choice allows us to have a compositional semantics which does not rely on any congruence rule (in particular the splitting rule used by [10]). Depending on the bookkeeping data structure used to instantiate the framework, we can obtain different causal semantics (i.e., [10, 6, 9]). We then show that our framework enjoys the standard properties for a reversible calculus, namely the loop lemma and causal consistency, regardless of the notion of causality which is used. We prove causal correspondence between the causal semantics introduced in [6] and the matching instance of our framework.

The rest of the paper is as follows: syntax and operational semantics of the framework are given in Section 2. In Section 3, we show how by using different data structures we can encompass different causal semantics. The main results are given in Section 4, and Section 5 concludes the paper. For reviewers' convenience full proofs are gathered in Appendix A.

## 2 The Framework

We present the syntax and operational semantics of our parametric framework, after an informal introduction.

## 2.1 Informal presentation

In [27] a general technique to reverse any CCS-like calculus is given. The key ideas are to use communication keys to identify events, and to make static all the operators of the calculus, since dynamic operators such as choice and prefix are *forgetful* operators. For example, if we take a CCS process $a.P \mid \overline{a}.Q$ a possible computation is:

$$a.P \mid \overline{a}.Q \xrightarrow{\tau[i]} a[i].P \mid \overline{a}[i].Q$$

As one can see, prefixes are not destroyed but *decorated* with a communication key. The obtained process acts like $P \mid Q$, since decorated prefixes are just used for backward steps. We bring this idea to the $\pi$-calculus. For example by lifting this process into $\pi$-calculus we have something like

$$a(x).P \mid \overline{a}b.Q \xrightarrow{\tau:i} a(x)[i].P\{b^i/x\} \mid \overline{a}b[i].Q$$

In the substitution $\{b^i/x\}$, name $b$ is decorated with the key $i$ to record that it was substituted for variable $x$ in the synchronisation identified by the communication key $i$. The key $i$ is also recorded in the memories.

By choosing to adapt the ideas of [27] to work with the $\pi$-calculus, we avoid using the splitting rule of $R\pi$ [10]. In $R\pi$ each process is monitored by a memory, $m \triangleright P$, which is in charge of recording all past events of the process. In this way, the past of the process is not recorded directly in the process. One drawback of this approach is that one needs to resort to a splitting rule of the form

$$m \triangleright (P \mid Q) \equiv \langle \uparrow \rangle \cdot m \triangleright P \mid \langle \uparrow \rangle \cdot m \triangleright Q$$

to let both $P$ and $Q$ execute. This rule is not associative and moreover, as shown in [20], introduces some undesired non-determinism, since equivalent processes performing the same action may become non-equivalent processes.

The framework has to remember extrusions, and in particular who was the extruder of a certain name, and what is the *contextual cause* for an action. For example in

$$\nu a(\overline{b}a \mid a(x).P) \xrightarrow{\overline{b}\langle\nu a\rangle:i} \nu a_{\{i\}}(\overline{b}a[i] \mid a(x).P) \xrightarrow{a(x):j} \nu a_{\{i\}}(\overline{b}a[i] \mid a(x)[j,i].P)$$

we have that after the extrusion, the restriction $\nu a$ does not disappear as in standard $\pi$-calculus, but remains where it was, becoming the memory $\nu a_{\{i\}}$ (introduced in [10]). This memory records the fact that name $a$ was extruded because of transition $i$. Moreover, since it is no longer a restriction but just a decoration, the following transition using name $a$ can take place. Transition $j$ uses $i$ as its contextual cause, indicating that the input action can happen on $a$ because it was extruded by $i$, and this is recorded in the process $a(x)[j,i].P$.

## 2.2 Syntax

We assume the existence of the following *denumerable* infinite mutually disjoint sets: the set $\mathcal{N}$ of names, the set $\mathcal{K}$ of keys, and the set $\mathcal{V}$ of variables. Moreover, we let $\mathcal{K}_* = \mathcal{K} \cup \{*\}$ where $*$ is a special key. We let $a,b,c$ range over $\mathcal{N}$; $x,y$ range over $\mathcal{V}$ and $i,j,k$ range over $\mathcal{K}$.

The syntax of the framework is depicted in Figure 1. *Processes*, given by the $P,Q$ productions, are the standard processes of the $\pi$-calculus [28]: **0** represents the idle process; $\overline{b}c.P$ is the *output*-prefixed process indicating the act of sending name $c$ over channel $b$; $b(x).P$ is the *input*-prefixed process indicating the act of receiving a value (which will be bound to the variable $x$) on channel $b$. Process $P \mid Q$ represents the *parallel* composition of two processes, while $\nu a(P)$ represents the fact that name $a$ is *restricted* in $P$.

Reversibility is defined on the top of the $\pi$-calculus. Unlike in the standard $\pi$-calculus, executed actions are not discarded. Each of them, followed by the memory, becomes a part of the process that we shall call the *history*. *Reversible processes* are given by $X,Y$ productions. A reversible process $\mathbf{P}$ is a

$$X, Y ::= \quad \mathbf{P} \mid \overline{b}^j a^{j_1}[i,K].X \mid b^j(x)[i,K].X \mid X \mid Y \mid \nu a_\Delta(X)$$
$$P, Q ::= \quad \mathbf{0} \mid \overline{b}c.P \mid b(x).P \mid P \mid Q \mid \nu a(P)$$

Figure 1: Syntax.

standard $\pi$-calculus process $P$ where channels are decorated with instantiators. As we shall see later on, instantiators are used to keep track of substitutions. In a prefix of the form $\overline{b}a$ or $b(x)$ we say that name $b$ is used in *subject position*, while name $a$ and variable $x$ are in *object position*. We shall use operators $sub(\cdot)$ and $obj(\cdot)$ to get respectively the subject and the object of a prefix. The prefix $\overline{b}^j a^{j_1}[i,K].X$ represents a *past output* recording the fact that in the past the process $X$ performed an output identified by key $i$ and that its contextual cause set was $K \subseteq \mathscr{K}_*$. Prefix $b^j(x)[i,K].X$ represents a *past input* recording the fact that the input was identified by key $i$ and its contextual cause set was $K$. If it is not relevant whether the prefix in the process is an input or an output, we shall denote it with $\alpha$ ($\alpha = \overline{b}^j a^{j_1}$ or $\alpha = b^j(x)$).

Following [10] the restriction operator $\nu a_\Delta$ is decorated with the memory $\Delta$ which keeps track of the extruders of a name $a$. As we shall see later on, we shall abstract away from the form of $\Delta$, as different data structures lead to different notions of causality. When $\mathtt{empty}(\Delta) = \mathit{true}$, the data structure is initialised and $\nu a_\Delta$ will act as the usual restriction operator $\nu a$ of the $\pi$-calculus. The set of reversible processes is denoted with $\mathscr{X}$.

To simplify manipulation with reversible processes, we shall define history and general context. History context represents the reversible process $X$ made of executed prefixes. For example, we can express the process $X = \overline{b}^* a^*[i,K].\overline{c}^* a^*[i',K'].\mathbf{P}$ as $X = \mathtt{H}[\mathbf{P}]$ with $\mathtt{H}[\bullet] = \overline{b}^* a^*[i,K].\overline{c}^* a^*[i',K'].\bullet$. General context is defined on the top of the history context by adding parallel and restriction operators on it. For example, the process $Z \mid Y \mid X$ can be written as $C[X]$ if the only relevant element is $X$. Formally:

**Definition 1** (History and General context). *History contexts* $\mathtt{H}$ *and general contexts* $C$ *are reversible processes with a hole* $\bullet$, *defined by the following grammar:*

$$\mathtt{H} ::= \bullet \mid \alpha[i,K].\bullet \qquad C ::= \mathtt{H}[\bullet] \mid X \mid \bullet \mid \nu a_\Delta(\bullet)$$

**Free names and free variables.** Notions of free names and free variables in our framework are standard. It suffices to note that constructs with binders are of the forms: $\nu a_\Delta(X)$ when $\mathtt{empty}(\Delta)$ holds, which binds the name $a$ with scope $X$; and $b(x).P$, which binds the variable $x$ with scope $P$. We denote with $\mathtt{fn}(P)$ and $\mathtt{fn}(X)$ the set of free names of $P$ and of $X$ respectively.

**Remark 1.** Annotation $b^*$ to a name $b$, used either in the subject or in the object position, indicates that name $b$ has no instantiators.

Since the framework will be parametric in the data structure $\Delta$, we specify it as an interface (in the style of a Java interface) by defining the operations that it has to offer.

**Definition 2.** $\Delta$ *is a data structure with the following defined operations:*

($i$) $\mathtt{init} : \Delta \to \Delta$ *initialises the data structure*

($ii$) $\mathtt{empty} : \Delta \to \mathtt{bool}$ *predicate telling whether* $\Delta$ *is empty*

($iii$) $+ : \Delta \times \mathscr{K} \to \Delta$ *operation adding a key to* $\Delta$

($iv$) $\#i : \Delta \times \mathscr{K} \to \Delta$ *operation removing a key from* $\Delta$

($v$) $\in : \Delta \times \mathscr{K} \to \mathtt{bool}$ *predicate telling whether a key belongs to* $\Delta$

We now define three instances of $\Delta$: sets, sets indexed with an element and sets indexed with a set. As we shall see, these three instances will give rise to three different notions of object causality.

**Set.** $\Gamma$ is a set containing keys (i.e. $\Gamma \subseteq \mathcal{K}$). The intuition of $va_\Gamma$ is that **any** of the elements contained in $\Gamma$ can be a contextual cause for $a$ (i.e., the reason why $a$ is known to the context).

**Definition 3** (Operations on a set). *The operations on a set $\Gamma$ are defined as:*

(*i*) $\texttt{init}(\Gamma) = \emptyset$

(*ii*) $\texttt{empty}(\Gamma) = \textit{true, when } \Gamma = \emptyset$

(*iii*) $+$ *is the classical addition of elements to a set*

(*iv*) *#i is defined as the identity, that is $\Delta_{\#i} = \Gamma_{\#i} = \Gamma$.*

(*v*) $i \in \Gamma$ *the key i belongs to the set $\Gamma$*

**Indexed set.** $\Gamma_w$ is an indexed set containing keys and $w$ is the key of the action which extruded a name $a$. In this case the contextual cause for name $a$ can be **just** $w$. If there is no cause, then we shall set $w = *$.

**Definition 4** (Operations on an indexed set). *The operations on an indexed set $\Gamma_w$ are defined as:*

(*i*) $\texttt{init}(\Gamma_w) = \emptyset_*$

(*ii*) $\texttt{empty}(\Gamma_w) = \textit{true, when } \Gamma = \emptyset \wedge w = *$

(*iii*) *operation $+$ is defined as:* $\Gamma_w + i = \begin{cases} (\Gamma \cup \{i\})_i, & \text{when } w = * \\ (\Gamma \cup \{i\})_w, & \text{when } w \neq * \end{cases}$

(*iv*) *operation #i is defined inductively as:*

$$(X \mid Y)_{\#i} = X_{\#i} \mid Y_{\#i} \qquad (\mathtt{H}[X])_{\#i} = \mathtt{H}[X_{\#i}] \qquad (\mathbf{P})_{\#i} = \mathbf{P}$$
$$(va_{\Gamma_i} X)_{\#i} = va_{\Gamma_*} X_{\#i} \qquad (va_{\Gamma_w} X)_{\#i} = va_{\Gamma_w} X_{\#i}$$

(*v*) $i \in \Gamma_w$ *the key i belongs to the set $\Gamma$, regardless of $w$ (e.g. $i \in \{i\}_*$)*

**Set indexed with a set.** $\Gamma_\Omega$ is a set containing keys indexed with a set $\Omega \in \mathcal{K}_*$. Extruders of name $a$ which are not part of the communication, will be saved in the set $\Omega$. In this case the contextual cause for name $a$ is a set $\Omega$. If there is no cause, then we shall set $\Omega = \{*\}$.

**Definition 5** (Operations on a set indexed with a set). *The operations on a set indexed with a set $\Gamma_\Omega$ are defined as:*

(*i*) $\texttt{init}(\Gamma_\Omega) = \emptyset_{\{*\}}$

(*ii*) $\texttt{empty}(\Gamma_\Omega) = \textit{true, when } \Gamma = \emptyset \wedge \Omega = \{*\}$

(*iii*) *operation $+$ is defined as:* $(\Gamma_\Omega) + i = (\Gamma \cup \{i\})_{(\Omega \cup \{i\})}$

(*iv*) *operation $_{\#i}$ is defined inductively as:*

$$(X \mid Y)_{\#i} = X_{\#i} \mid Y_{\#i} \qquad (va_{\Gamma_\Omega} X)_{\#i} = va_{\Gamma_{\Omega \setminus \{i\}}} X_{\#i} \qquad (\mathtt{H}(X))_{\#i} = \mathtt{H}[X_{\#i}] \qquad (\mathbf{P})_{\#i} = \mathbf{P}$$

(*v*) $i \in \Gamma_\Omega$ *the key i belongs to the set $\Gamma$, regardless $\Omega$ (e.g. $i \in \{i\}_{\{*\}}$)*

$$(\text{OUT1}) \; \overline{b}^j a^{j_1}.P \xrightarrow{(i,K,j):\overline{b}a} \overline{b}^j a^{j_1}[i,K].P \qquad (\text{OUT2}) \; \dfrac{X \xrightarrow{(i,K,j):\overline{b}a} X' \quad \texttt{fresh}(i,\texttt{H}[X])}{\texttt{H}[X] \xrightarrow{(i,K,j):\overline{b}a} \texttt{H}[X']}$$

$$(\text{IN1}) \; b^j(x).P \xrightarrow{(i,K,j):b(x)} b^j(x)[i,K].P \qquad (\text{IN2}) \; \dfrac{X \xrightarrow{(i,K,j):b(x)} X' \quad \texttt{fresh}(i,\texttt{H}[X])}{\texttt{H}[X] \xrightarrow{(i,K,j):b(x)} \texttt{H}[X']}$$

$$(\text{PAR}) \; \dfrac{X \xrightarrow{(i,K,j):\pi} X' \quad i \notin Y}{X \mid Y \xrightarrow{(i,K,j):\pi} X' \mid Y} \qquad (\text{RES}) \; \dfrac{X \xrightarrow{(i,K,j):\pi} X' \quad a \notin \pi}{\nu a_\Delta(X) \xrightarrow{(i,K,j):\pi} \nu a_\Delta(X')}$$

$$(\text{COM}) \; \dfrac{X \xrightarrow{(i,K,j):\overline{b}a} X' \quad Y \xrightarrow{(i,K',j'):b(x)} Y' \quad K =_* j' \wedge K' =_* j}{X \mid Y \xrightarrow{(i,*,*):\tau} X' \mid Y'\{a^i/x\}}$$

Figure 2: Rules that are common to all instances of the framework.

## 2.3 Operational Semantics

The grammar of the labels generated by the framework is:

$$\mu ::= (i,K,j) : \pi \qquad \pi ::= \overline{b}c \mid b(x) \mid \overline{b}\langle \nu c_\Delta \rangle \mid \tau$$

where $i$ is the key, and $K \subseteq \mathcal{K}_*$, $j \in \mathcal{K}_*$ are the set of contextual causes and an instantiator of $i$, respectively. If there is no action which caused and/or instantiated $i$, we denote this with $K = \{*\}$, $j = *$, respectively. The set $\mathscr{L}$ of all possible labels generated by the framework is defined as $\mathscr{L} = \mathcal{K} \times \mathcal{K}_* \times \mathcal{K}_* \times \mathscr{A}$, where $\mathscr{A}$ is a set of actions ranged over by $\pi$. We extend $sub(\cdot)$ and $obj(\cdot)$ to apply also to labels.

The operational semantics of the reversible framework is given in terms of a labelled transition system (LTS) $(\mathscr{X}, \mathscr{L}, \rightarrow)$, where $\mathscr{X}$ is the set of reversible processes; $\rightarrow = \twoheadrightarrow \cup \rightsquigarrow$ where $\twoheadrightarrow$ is the least transition relation induced by the rules in Figures 2 and 3; and $\rightsquigarrow$ is the least transition relation induced by the rules in Figure 4.

**Definition 6** (Process keys). *The set of* communication keys *of a process $X$, written $\texttt{key}(X)$, is inductively defined as follows:*

$$\texttt{key}(X \mid Y) = \texttt{key}(X) \cup \texttt{key}(Y) \qquad \texttt{key}(\alpha[i,K].X) = \{i\} \cup \texttt{key}(X)$$
$$\texttt{key}(\nu a_\Delta(X)) = \texttt{key}(X) \qquad \texttt{key}(P) = \emptyset$$

**Definition 7.** *A key $i$ is* fresh *in a process $X$, written $\texttt{fresh}(i,X)$ if $i \notin \texttt{key}(X)$.*

The forward rules of a framework are divided into two *groups*, depending on whether they are parametric with respect to $\Delta$ or they are common to all the instances of the framework.

Common rules are given in Figure 2. Rules OUT1 and IN1 generate a fresh new key $i$ which is bound to the action. Rules OUT2 and IN2 inductively allow a prefixed process $\texttt{H}[X]$ to execute if $X$ can execute. Condition $i \notin Y$ in rule PAR ensures that action keys are unique. Rule RES is defined in the usual way. Two processes can synchronise through the rule COM if the additional condition is satisfied ($K =_* j$ means $* \in K$ or $j = *$ or $K = j$). After the communication, necessary substitution is applied to the rest of the input process. In the process $Y'\{a^i/x\}$ every occurrence of variable $x \in \texttt{fn}(Y')$ is substituted with the name $a^i$, that is, the name $a$ decorated with the key $i$ of the action which was executed. In the further

$$(\text{Cause Ref}) \ \frac{X \xrightarrow{(i,K,j):\pi} X' \quad a \in sub(\pi) \quad \texttt{empty}(\Delta) \neq true \quad \texttt{Cause}(\Delta, K, K')}{va_\Delta(X) \xrightarrow{(i,K',j):\pi} va_\Delta(X'_{[K'/K]\&i})}$$

$$(\text{Open}) \ \frac{X \xrightarrow{(i,K,j):\pi} X' \quad \pi = \overline{b}a \vee \pi = \overline{b}\langle va_{\Delta'}\rangle \quad \texttt{Update}(\Delta, K, K')}{va_\Delta(X) \xrightarrow{(i,K',j):\overline{b}\langle va_\Delta\rangle} va_{\Delta+i}(X'_{[K'/K]\&i})}$$

$$(\text{Close}) \ \frac{X \xrightarrow{(i,K,j):\overline{b}\langle va_\Delta\rangle} X' \quad Y \xrightarrow{(i,K',j'):b(x)} Y' \quad K =_* j' \wedge K' =_* j}{X \mid Y \xrightarrow{(i,*,*):\tau} va_\Delta(X'_{\#i} \mid Y'\{a^i/x\})}$$

Figure 3: Parametric rules

actions of a process $Y'\{a^i/x\}$, the key $i$ will be called the *instantiator*. The instantiators are used just to keep track of the substitution, not to define a name. For example, the two processes $\overline{b}^j a^*.\mathbf{P}$ and $b^{j'}(x).\mathbf{P}'$ can communicate, even if the instantiators of the name $b$ are not the same. Let us note that we use a *late* semantics, since substitution happens in the rule COM. In order to understand how the basic rules work let us consider the following example.

**Example 1.** *Let* $X = \overline{b}^* a^*.\mathbf{0} \mid b^*(x).\overline{x}c^*$. *There are two possibilities for the process X:*

- *process X can preform an output and an input action on the channel b while synchronising with environment:*

$$\overline{b}^* a^*.\mathbf{0} \mid b^*(x).\overline{x}c^* \xrightarrow{(i,*,*):\overline{b}a} \overline{b}^* a^*[i,*].\mathbf{0} \mid b^*(x).\overline{x}c^* \xrightarrow{(i',*,*):b(x)} \overline{b}^* a^*[i,*].\mathbf{0} \mid b^*(x)[i',*].\overline{x}c^* = Y_1$$

  *As we can notice, the output action $\overline{b}a$ is identified by key $i$, while the input action is identified by key $i'$.*

- *The synchronisation can happen inside of the process X:*

$$\overline{b}^* a^*.\mathbf{0} \mid b^*(x).\overline{x}c^* \xrightarrow{(i,*,*):\tau} \overline{b}^* a^*[i,*].\mathbf{0} \mid b^*(x)[i,*].\overline{a}^i c^* = Y_2$$

  *We can notice that $\tau$ action is identified with key $i$ and during the synchronisation variable $x$ is substituted with a received name a decorated with the key $i$ of the executed action. In this way we keep track of the substitution of a name.*

We now define the operation $X_{[K'/K]\&i}$, which updates the contextual cause $K$ of an action identified by $i$ with the new cause $K'$. Contextual cause update will be used in the the parametric rules of Figure 3 (OPEN and CAUSE REF). Formally:

**Definition 8** (Contextual Cause Update). *The* contextual cause update *of a process, written $X_{[K'/K]\&i}$ is defined as follows:*

$$(X \mid Y)_{[K'/K]\&i} = X_{[K'/K]\&i} \mid Y_{[K'/K]\&i} \qquad \texttt{H}[\alpha[i,K].X]_{[K'/K]\&i} = \texttt{H}[\alpha[i,K'].X]$$
$$(va_\Delta(X))_{[K'/K]\&i} = va_\Delta(X)_{[K'/K]\&i} \qquad \texttt{H}[\alpha[j,K].X]_{[K'/K]\&i} = \texttt{H}[\alpha[j,K].X]$$

Parametric rules are given in Figure 3. Depending on the underlying causal semantics the way a contextual cause is chosen differs. This is why we need to define two predicates: $\texttt{Cause}(\cdot)$ and $\texttt{Update}(\cdot)$.

$$(\text{Out1}^\bullet) \ \overline{b}^j a[i,K].P \xrightarrow{(i,K,j):\overline{b}a} \overline{b}^j a.P \qquad (\text{Out2}^\bullet) \ \frac{X' \xrightarrow{(i,K,j):\overline{b}a} X \quad \texttt{fresh}(i,\texttt{H}[X])}{\texttt{H}[X'] \xrightarrow{(i,K,j):\overline{b}a} \texttt{H}[X]}$$

$$(\text{In1}^\bullet) \ b^j(x)[i,K].P \xrightarrow{(i,K,j):b(x)} b^j(x).P \qquad (\text{In2}^\bullet) \ \frac{X' \xrightarrow{(i,K,j):b(x)} X \quad \texttt{fresh}(i,\texttt{H}[X])}{\texttt{H}[X'] \xrightarrow{(i,K,j):b(x)} \texttt{H}[X]}$$

$$(\text{Par}^\bullet) \ \frac{X' \xrightarrow{(i,K,j):\pi} X \quad i \notin Y}{X' \mid Y \xrightarrow{(i,K,j):\pi} X \mid Y} \qquad (\text{Com}^\bullet) \ \frac{X' \xrightarrow{(i,K,j):\overline{b}a} X \quad Y' \xrightarrow{(i,K',j'):b(x)} Y \quad K =_* j' \wedge K' =_* j}{X' \mid Y' \xrightarrow{(i,*,*):\tau} X \mid Y\{x/a^i\}}$$

$$(\text{Res}^\bullet) \ \frac{X' \xrightarrow{(i,K,j):\pi} X \quad a \notin \pi}{va_\Delta(X') \xrightarrow{(i,K,j):\pi} va_\Delta(X)} \qquad (\text{Open}^\bullet) \ \frac{X' \xrightarrow{(i,K,j):\pi} X \quad \pi = \overline{b}a \vee \pi = \overline{b}\langle va_{\Delta'}\rangle \quad \texttt{Update}(\Delta,K,K')}{va_{\Delta+i}(X') \xrightarrow{(i,K',j):\overline{b}\langle va_\Delta\rangle} va_\Delta(X)}$$

$$(\text{Cause Ref}^\bullet) \ \frac{X' \xrightarrow{(i,K,j):\pi} X \quad a \in sub(\pi) \quad \texttt{empty}(\Delta) \neq true \quad \texttt{Cause}(\Delta,K,K')}{va_\Delta(X') \xrightarrow{(i,K',j):\pi} va_\Delta(X)}$$

$$(\text{Close}^\bullet) \ \frac{X' \xrightarrow{(i,K,j):\overline{b}\langle va_\Delta\rangle} X \quad Y' \xrightarrow{(i,K',j'):b(x)} Y \quad K =_* j' \wedge K' =_* j}{va_\Delta(X' \mid Y') \xrightarrow{(i,*,*):\tau} X \mid Y\{x/a^i\}}$$

Figure 4: Backward rules.

When instantiating $\Delta$ with a specific data structure, different implementations of such predicates are needed. We shall define them precisely when discussing various causal semantics in the Section 3. Every time a label produced by a process $X$ passes the restriction $va_\Delta$ it needs to check if it is necessary to modify the contextual cause. Depending on whether name $a$ is in the subject or in the object position in the label of an action, rules CAUSE REF or OPEN can be used, respectively. Rule CAUSE REF is used when the subject of a label is an already extruded name and a predicate $\texttt{Cause}(\Delta,K,K')$ tells whether contextual cause $K$ has to be substituted with $K'$. Rule OPEN deals with the scope extrusion of a restricted name. If the restricted name $a$ is used as object of a label with key $i$ we have to record that $i$ is one of the potential extruders of $a$. Naturally, if $\texttt{empty}(\Delta) = true$ then the first extruder initialises the data structure. Also in this case it might happen that we have to update the contextual cause of the label $i$. This is why predicate $\texttt{Update}(\Delta,K,K')$ is used. Two processes can synchronise through the rule CLOSE satisfying the additional condition. In some semantics, silent actions do not bring the causal information on what is a reason to introduce the operator $\#_i$, where every time when an extruded name is closed over the context, the key of the closing action is deleted from indexes of $\Delta_h$ in all restrictions $va_{\Delta_h} \in X'$.

Backward rules are symmetric to the forward ones; they are presented in Figure 4. The predicates are not necessary for the backward transitions as they are invariant in the history of processes but we keep them to simplify the proofs. In order to better understand the backward rules, we shall consider the following example.

**Example 2.** *Let us consider the following processes from Example 1:*

- $Y_1 = \overline{b}^* a^*[i,*].\mathbf{0} \mid b^*(x)[i',*].\overline{x}c^*$; *Process $Y_1$ can perform backward actions on the channel b (an* *input action identified with key $i'$ and an output action identified with key i) in any order. For example, let us revers first the input and then the output action:*

$$Y_1 = \overline{b}^* a^*[i,*].\mathbf{0} \mid b^*(x)[i',*].\overline{x}c^* \xrightsquigarrow{(i',*,*):b(x)} \overline{b}^* a^*[i,*].\mathbf{0} \mid b^*(x).\overline{x}c^* \xrightsquigarrow{(i,*,*):\overline{b}a} \overline{b}^* a^*.\mathbf{0} \mid b^*(x).\overline{x}c^* = X$$

*We shall notice that all the necessary elements to reverse the action $b(x)$ are saved in the history part of the process $Y_1$.*

- $Y_2 = \overline{b}^* a^*[i,*].\mathbf{0} \mid b^*(x)[i,*].\overline{a}^i c^*$; *Process $Y_2$ can reverse the communication happened on the channel b, between its subprocesses. Due to side condition of the rule* PAR$^\bullet$, *it is impossible to reverse an input or an output action separately:*

$$\overline{b}^* a^*[i,*].\mathbf{0} \mid b^*(x)[i,*].\overline{a}^i c^* \not\xrightsquigarrow{(i,*,*):\overline{b}a} \overline{b}^* a^*.\mathbf{0} \mid b^*(x)[i,*].\overline{a}^i c^*$$

*The backward action above cannot be executed as the key i belongs to the process in parallel ($i \in$ key$(b^*(x)[i,*].\overline{a}^i c^*)$). The only possible backward step is:*

$$\overline{b}^* a^*[i,*].\mathbf{0} \mid b^*(x)[i,*].\overline{a}^i c^* \xrightsquigarrow{(i,*,*):\tau} \overline{b}^* a^*.\mathbf{0} \mid b^*(x).\overline{x}c^* = X$$

**Remark 2.** The choice operator $(+)$, can be easily added to the framework by following the approach of [27] and by making the operator static.

## 3 Mapping causal semantics of $\pi$ into the framework

We now review three notions of causal semantics for $\pi$-calculus and show how to map them into our framework by giving the definitions for the side conditions in the rules in Figure 3.

*R$\pi$-calculus.* Cristescu et al [10] introduce a compositional semantics for the reversible $\pi$-calculus. Information about the past actions is kept in a memory added to every process. A term of the form $m \triangleright P$ represents a reversible process, where memory $m$ is a stack of events and $P$ is the process itself. A memory contains two types of events, one which keeps track of the past action, $\langle i, k, \pi \rangle$, where elements of a triple are the key, the contextual cause and the executed action, respectively; and one which keeps track of the position of the process in the parallel composition, $\langle \uparrow \rangle$. Before executing in parallel, a process splits by duplicating its memory and adding event $\langle \uparrow \rangle$ on the top of each copy. This is achievable with specially defined structural congruence rules. The use in [10] of indexed restriction $\nu a_\Gamma$ was the inspiration for our parametric indexed restriction $\nu a_\Delta$. A key point of the semantics of [10] is that it enjoys certain correctness properties: one of which is that two visible transitions are causally related iff for all contexts the corresponding silent transitions are. Since an action can be caused only through the *subject* of a label we have that contextual cause $K$ will be a singleton. We shall consider one relation between the prefixes into the history. In this way, while changing the cause with the rule CAUSE REF, the condition Cause($\cdot$) needs to keep track of the instantiation of the cause.

**Definition 9** (Instantiation relation). *Two keys $i_1$ and $i_2$ such that $i_1, i_2 \in$ key$(X)$ and $X = C[b^*(x)[i_1, K_1].Y]$ with $Y = C'[\alpha^{j_2}[i_2, K_2].Z]$, are in instantiation relation, $i_1 \rightsquigarrow_X i_2$, if $j_2 = i_1$. If $i_1 \rightsquigarrow_X i_2$ holds, we will write $K_1 \rightsquigarrow_X K_2$.*

To obtain $R\pi$ causality in our framework, we need to instantiate the rules of Figure 3 with the following predicates.

**Definition 10** ($R\pi$ causality)**.** *If data structure $\Delta$ is instantiated with a set $\Gamma$, the predicates from Figure 3 are defined as:*

1. $\texttt{Cause}(\Delta, K, K') = \texttt{Cause}(\Gamma, K, K')$ *stands for $K = K'$ or $\exists K' \in \Gamma \; K \rightsquigarrow_X K'$;*

2. $\texttt{Update}(\Delta, K, K') = \texttt{Update}(\Gamma, K, K')$ *stands for $K' = K$.*

The predicates defined above coincide with the conditions of the semantics introduced in [10]. In the following example we shall give the intuition of the $R\pi$ causality using our framework.

**Example 3.** *Let us consider the process $X = \nu a_\emptyset(\overline{b}^* a^* \mid \overline{c}^* a^* \mid a^*(x))$. By applying rule* OPEN *twice and executing concurrently two extrusions on names b and c, we obtain a process:*

$$\nu a_{\{i,h\}}(\overline{b}^* a^*[i,*] \mid \overline{c}^* a^*[h,*] \mid a^*(x))$$

*The rule* CAUSE REF *is used for the execution of the third action. By definition of the predicate* $\texttt{Cause}(\cdot)$, *the action $a(x)$ can choose its cause from a set $\{i,h\}$. By choosing h for example, and executing the input action, we get the process:*

$$\nu a_{\{i,h\}}(\overline{b}^* a[i,*] \mid \overline{c}^* a[h,*] \mid a^*(x)[l,h])$$

*In the memory $[l,h]$ we can see that the action identified with key $l$ needs to be reversed before the action with key $h$. Process $\overline{b}^* a[i,*]$ can execute a backward step at any time with the rule* OPEN$^\bullet$.

**Boreale-Sangiorgi and Degano-Priami causal semantics.** A compositional causal semantics for standard (i.e., forward only) $\pi$-calculus was introduced by Boreale and Sangiorgi [6]. Later on, Degano and Priami in [15] introduced a causal semantics for $\pi$ based on localities. While using different approaches to keep track of the dependences in $\pi$-calculus, these two approaches impose the same order of the forward actions (as claimed in [15]). Hence, from the reversible point of view we can take it that the causality notions of these two semantics coincide. In what follows we shall concentrate on the Boreale-Sangiorgi causal semantics. To show the correspondence between the mentioned semantics and our framework, we shall consider it in a late (rather than early, as originally given) version. The precise definition is given in Appendix A.3.

The authors distinguish between two types of causality: *subject* and the *object*. To capture the first one, they introduce a causal term $K :: A$, where $K$ is a set of causes recording that every action performed by $A$ depends on $K$. The object causality is defined on the run (trace) of a process in such a way that once a bound name been extruded, it causes all the subsequent actions using that name in any position of the label. Since an action can be caused through the *subject* and *object* position of a label, the contextual cause is a set $K \subseteq \mathscr{K}_*$. For example, let us consider a process $\nu a(\nu b(\overline{c}b \mid \overline{d}a \mid \overline{b}a))$ with a trace $\xrightarrow{\overline{c}\langle \nu b \rangle} \xrightarrow{\overline{d}\langle \nu a \rangle} \xrightarrow{\overline{b}a}$. The action $\overline{b}a$ depends on the first action because with it name $b$ was extruded and on the second action because with it name $a$ was extruded. It is important to remark that a silent action does not exhibit causes.

To capture Boreale-Sangiorgi late semantics we need to give definitions for the predicates in Figure 3.

**Definition 11** (Boreale-Sangiorgi causal semantics)**.** *If an indexed set $\Gamma_w$ is chosen as a data structure for a memory $\Delta$, the predicates from Figure 3 are defined as:*

1. $\texttt{Cause}(\Delta, K, K') = \texttt{Cause}(\Gamma_w, K, K')$ *stands for $K' = K \cup \{w\}$*

2. $\texttt{Update}(\Delta, K, K') = \texttt{Update}(\Gamma_w, K, K')$ *stands for $K' = K \cup \{w\}$*

Let us comment on the above definition. After the first extrusion of a name, the cause is fixed and there is no possibility of choosing another cause from the set $\Gamma$. To capture this behaviour we use the key of the first extruder, say $w$, as the index of the set $\Gamma$. The following example explains how our framework captures Boreale-Sangiorgi causality. We shall use the same process as in Example 3.

**Example 4.** *Consider the process* $X = \nu a_{\emptyset_*}(\overline{b}^* a^* \mid \overline{c}^* a^* \mid a^*(x))$. *By applying rule* OPEN *and executing the first extrusion on name b, we obtain the process:*

$$\nu a_{\{i\}_i}(\overline{b}^* a^*[i, *] \mid \overline{c}^* a^* \mid a^*(x))$$

*In the memory* $\{i\}_i$ *the index i indicates that name a was extruded with the action i. On the process* $\overline{c}^* a^*$, *rule* OPEN *can be applied. By definition of the predicate* Update$(\cdot)$, *the output action is forced to add* $w = i$ *in its cause set. Similar for the process* $a^*(x)$, *by applying the rule* CAUSE REF *and definition of the predicate* Cause$(\cdot)$. *After two executions, we obtain the process:*

$$\nu a_{\{i,h\}_i}(\overline{b}^* a^*[i, *] \mid \overline{c}^* a^*[h, \{i, *\}] \mid a^*(x)[l, \{i, *\}])$$

*In the memories* $[h, \{i, *\}]$ *and* $[l, \{i, *\}]$ *we see that both executed actions are caused by action i and this is why it needs to be reversed last. The second and the third action can be reversed in any order.*

**Crafa, Varacca and Yoshida causal semantics.** The authors introduced a compositional event structure semantics for the forward $\pi$-calculus [9]. They represent a process as a pair $(E, \mathtt{X})$, where $E$ is a prime event structure and $\mathtt{X}$ is a set of bound names. Disjunctive objective causality is represented in such a way that an action with extruded name in the subject position can happen if at least one extrusion of that name has been executed before. In the case of parallel extrusions of the same name, an action can be caused by any of them, but it is not necessary to remember which one.

Consequently, events do not have a unique causal history. As discussed in [11] this type of disjunctive causality cannot be expressed when we consider processes with a contexts. To adapt this notion of causality to reversible settings we need to keep track of causes; otherwise by going backwards we could reach an undefined state (where the extruder of a name is reversed, but not the action using that name in the subject position).

We consider two possibilities for keeping track of causes: the first one is by choosing one of the possible extruders and the second one is recording all of them. In the first case, we would obtain a notion of causality similar to the one introduced in [10]. In the following we shall concentrate on the second option. The idea is that, since we do not know which extruder really caused the action on an extruded name, we shall record the whole set of extruders that happened previously. In the framework, the set of executed extruders is set $\Omega$. The extrusions which are part of synchronisations will be deleted from $\Omega$ with the operation #.

The predicates from the rules of Figure 3 are defined as follows:

**Definition 12** (Disjunctive causality). *If an indexed set $\Gamma_\Omega$ is chosen as a data structure for a memory $\Delta$, the predicates are defined as:*

1. Cause$(\Delta, K, K') =$ Cause$(\Gamma_\Omega, K, K')$ *stands for* $K' = K \cup \Omega$

2. Update$(\Delta, K, K') =$ Update$(\Gamma_\Omega, K, K')$ *stands for* $K' = K$

In the following example we shall give the intuition of how our framework captures the defined notion of causality.

**Example 5.** *Let us consider the process $X = va_{\emptyset_{\{*\}}}(\overline{b}^*a^* \mid \overline{c}^*a^* \mid a^*(x))$. By applying a rule* OPEN *twice and executing concurrently two extrusions on names b and c, we obtain a process:*

$$va_{\{i,h\}_{\{*,i,h\}}}(\overline{b}^*a^*[i,*] \mid \overline{c}^*a^*[h,*] \mid a^*(x))$$

*By definition of the predicate* Cause$(\cdot)$*, the third action will take the whole index set $\{*,i,h\}$ as a set K and we get the process:*

$$va_{\{i,h\}_{\{*,i,h\}}}(\overline{b}^*a[i,*] \mid \overline{c}^*a[h,*] \mid a^*(x)[l,\{*,i,h\}])$$

*In the memory $[l,\{*,i,h\}]$ we see that the first action to be reversed is the action with key l; the other two actions can be reversed in any order.*

## 4 Properties

In this section we shall show some properties of our framework. First we shall show that the framework is a conservative extension of standard π-calculus and that it enjoys causal consistency, a fundamental property for reversible calculi. After that, we shall prove causal correspondence between the causality induced by Boreale-Sangiorgi semantics and causality in the framework when $\Delta = \Gamma_w$. All proofs are gathered in Appendix A.

**Definition 13** (Initial and Reachable process). *A reversible process X is* initial *if it is derived from a π-calculus process P where all the restricting operators are initialised and in every prefix, names are decorated with a $*$. A reversible process is* reachable *if it can be derived from an initial process by using the rules in Figures 2, 3 and 4.*

### 4.1 Correspondence with the π-calculus

We now show that our framework is a conservative extension of the π-calculus. To do so, we first define an erasing function $\varphi$ that given a reversible process $X$, by deleting all the past information, generates a π process. Then we shall show that there is a *forward* operational correspondence between a reversible process $X$ and $\varphi(X)$. Let $\mathscr{P}$ be the set of π-calculus processes; then we have:

**Definition 14** (Erasing function). *The function $\varphi : \mathscr{X} \to \mathscr{P}$ that maps reversible processes to the π-calculus, is inductively defined as follows:*

$$\varphi(X \mid Y) = \varphi(X) \mid \varphi(Y) \qquad \varphi(H[X]) = \varphi(X) \qquad \varphi(0) = 0$$
$$\varphi(va_\Delta(X)) = \varphi(X) \quad \text{if } \mathtt{empty}(\Delta) = \mathit{false} \qquad \varphi(b^j(x).P) = b(x).\varphi(P)$$
$$\varphi(va_\Delta(X)) = va\ \varphi(X) \quad \text{if } \mathtt{empty}(\Delta) = \mathit{true} \qquad \varphi(\overline{b}^j a^{j'}.P) = \overline{b}a.\varphi(P)$$

*The erasing function can be extended to labels as:*

$$\varphi((i,K,j):\pi) = \varphi(\pi) \qquad\qquad \varphi(\overline{b}a) = \overline{b}a$$
$$\varphi(\overline{b}\langle va_\Delta\rangle) = \overline{b}\langle va\rangle \quad \text{when } \mathtt{empty}(\Delta) = \mathit{true} \qquad \varphi(b(x)) = b(x)$$
$$\varphi(\overline{b}\langle va_\Delta\rangle) = \overline{b}a \quad \text{when } \mathtt{empty}(\Delta) = \mathit{false} \qquad \varphi(\tau) = \tau$$

As expected the erasing function discards the past prefixes and name restriction operators that are non-empty. Moreover, it deletes all the information about the instantiators.

Every forward move of a reversible process $X$ can be matched in the π-calculus. To this end we use $\twoheadrightarrow_\pi$ to indicate the transition semantics of the π-calculus.

**Lemma 1.** *If there is a transition $X \xrightarrow{\mu} Y$ then $\varphi(X) \xrightarrow{\varphi(\mu)}_\pi \varphi(Y)$.*

We can state the converse of Lemma 1 as follows:

**Lemma 2.** *If there is a transition $P \xrightarrow{\varphi(\mu)}_\pi Q$ then for all reachable $X$ such that $\varphi(X) = P$, there is a transition $X \xrightarrow{\mu} Y$ with $\varphi(Y) = Q$.*

**Corollary 1.** *The relation given by $(X, \varphi(X))$, for all reachable processes $X$, is a strong bisimulation.*

## 4.2 The main properties of the framework

We now prove some properties of our framework which are typical of a reversible process calculus [12, 27, 20, 10]. Most of the terminology and the proof schemas are adapted from [12, 10] with more complex arguments due to the generality of our framework. The first important property is the so-called Loop Lemma, stating that any reduction step can be undone. Formally:

**Lemma 3** (Loop Lemma). *For every reachable process $X$ and forward transition $t : X \xrightarrow{\mu} Y$ there exists a backward transition $t' : Y \xrightarrow{\mu} X$, and conversely.*

Before stating our main theorems we need to define the causality relation. It is defined on the general framework and it is interpreted as the union of *structural* and *object* causality.

**Definition 15** (Structural cause on the keys). *For every two keys $i_1$ and $i_2$ such that $i_1, i_2 \in \text{key}(X)$, we let $i_1 \sqsubset_X i_2$ if $X = C[\alpha[i_1, K_1].Y]$ and $i_2 \in \text{key}(Y)$.*

Once having defined structural causal relation on keys, we can extend it to transitions.

**Definition 16** (Structural causality). *Transition $t_1 : X \xrightarrow{(i_1,K_1,j_1):\pi_1} X'$ is a structural cause of transition $t_2 : X'' \xrightarrow{(i_2,K_2,j_2):\pi_2} X'''$, written $t_1 \sqsubset t_2$, if $i_1 \sqsubset_{X'''} i_2$ or $i_2 \sqsubset_X i_1$. Structural causality, denoted with $\sqsubseteq$, is the reflexive and transitive closure of $\sqsubset$.*

Object causality is defined directly on the transitions and to keep track of it we use the contextual cause $K$.

**Definition 17** (Reverse transition). *The reverse transition of a transition $t : X \xrightarrow{\mu} Y$, written $t^\bullet$, is the transition with the same label and the opposite direction $t^\bullet : Y \xrightarrow{\mu} X$, and vice versa. Thus $(t^\bullet)^\bullet = t$.*

**Definition 18** (Object causality). *Transition $t_1 : X \xrightarrow{(i_1,K_1,j_1):\pi_1} X'$ is an object cause of transition $t_2 : X' \xrightarrow{(i_2,K_2,j_2):\pi_2} X''$, written $t_1 < t_2$, if $i_1 \in K_2$ or $i_2 \in K_1$ (for the backward transition) and $t_1 \neq t_2^\bullet$. Object causality, denoted with $\ll$, is the reflexive and transitive closure of $<$.*

**Example 6.** *Consider a process $X = \nu a_\Delta(\overline{b}^* a^* \mid \overline{c}^* a^* \mid a^*(z))$ and the case when $\Delta = \emptyset_*$, as in Example 4. The executed actions would be $\xrightarrow{(i_1,*,*):\overline{b}\langle \nu a_{\emptyset_*}\rangle} \xrightarrow{(i_2,\{i_1\},*):\overline{c}\langle \nu a_{\{i_1\}_{i_1}}\rangle} \xrightarrow{(i_3,\{i_1\},*):a(z)}$. We can notice that $K_2 = \{i_1\}$ and $K_3 = \{i_1\}$, indicating that the second and the third action are caused by the first one. By choosing a different data structure we can obtain different causal order, as mentioned in Example 3 and Example 5.*

**Definition 19** (Causality relation and concurrency). *The causality relation $\prec$ is the reflexive and transitive closure of structural and object cause: $\prec = (\sqsubseteq \cup \ll)^*$. Two transitions are concurrent if they are not causally related.*

Concurrent transitions can be permuted, and the commutation of transitions is preserved up to label equivalence.

**Definition 20** (Label equivalence). *Label equivalence, $=_\lambda$, is the least equivalence relation satisfying: $(i,K,j) : \bar{b}\langle va_\Delta \rangle =_\lambda (i,K,j) : \bar{b}\langle va_{\Delta'} \rangle$ for all $i,j,K,a,b$ and $\Delta, \Delta' \subseteq \mathcal{K}$. (Having an indexed set $\Gamma_w$ for $\Delta$ we disregard index $w$, and observe $\Gamma \subseteq \mathcal{K}$.)*

**Lemma 4** (Square Lemma). *If $t_1 : X \xrightarrow{\mu_1} Y$ and $t_2 : Y \xrightarrow{\mu_2} Z$ are two concurrent transitions, there exist $t_2' : X \xrightarrow{\mu_2'} Y_1$ and $t_1' : Y_1 \xrightarrow{\mu_1'} Z$ where $\mu_i =_\lambda \mu_i'$.*

We shall follow the standard notation and say that $t_2$ is a residual of $t_2'$ after $t_1$, denoted with $t_2 = t_2'/t_1$. Two transitions are *coinitial* if they have the same source, *cofinal* if they have the same target, and *composable* if the target of one is the source of the other. A sequence of pairwise composable transitions is called a *trace*, written as $t_1; t_2$. We denote with $\varepsilon$ the *empty* trace. Notions of target, source, composability and reverse extend naturally to traces.

With the next theorem we prove that reversibility in our framework is causally consistent.

**Definition 21** (Equivalence up-to permutation). *Equivalence up-to permutation, $\sim$, is the least equivalence relation on the traces, satisfying:*

$$t_1;(t_2/t_1) \sim t_2;(t_1/t_2) \qquad t;t^\bullet \sim \varepsilon$$

Equivalence up-to permutation introduced in [12] is an adaptation of equivalence between traces introduced in [21, 7] that additionally erases from a trace, transitions triggered in both directions. It just states that concurrent actions can be swapped and that a trace made by a transition followed by its inverse is equivalent to the empty trace.

**Theorem 1.** *Two traces are coinitial and cofinal if and only if they are equivalent up-to permutation.*

### 4.3 Correspondence with Boreale and Sangiorgi's semantics

We prove causal correspondence between Boreale and Sangiorgi's late semantics (rather than early, as originally given) and the framework when memory $\Delta$ is instantiated with $\Gamma_w$. The precise definitions and the proofs are given in Appendix A.3, here we shall give just briefly presentation of the idea.

To compare semantics, we observe traces (runs) of the processes. Labels in the framework will bring additional information about the multiset of the structural causes ($K_F$) of the executed action and a trace in the framework will have the following form: $X_1 \xrightarrow[K_{F1}]{\mu_1} X_2 \ldots X_{n-1} \xrightarrow[K_{Fn}]{\mu_n} X_n$. To simplify notation, we shall write the transition $A \xrightarrow[K;k]{\pi} A_2$ from Boreale and Sangiorgi's semantics as $A \xrightarrow[K_B]{\zeta} A_2$, where $\zeta = k : \pi$.

Focusing on the structural correspondence, the main difference is in the silent actions. In the framework, silent actions are identified with unique keys, while in Boreale and Sangiorgi's semantics, they just merge the cause sets of the actions participating in the communication. Hence, we need to provide a connection between sets of structural causes in these two semantics. Let us briefly explain our method; more details can be found in Appendix A.3.

Suppose that we have two transition $t$ and $t'$, where $t : X \xrightarrow[K_F]{(i,K,j):\pi} X'$ and $t' : A \xrightarrow[K_B]{i:\pi} A'$ where the continuation of the processes $X$ and $A$ is $\pi.P$[1]. We can represent the dependences between the keys in the history of the process $X$ (all executed actions in $X$) with a directed graph, in the following way: keys of executed actions will be represented as vertices of a graph (actions which are part of a communication and have the same key, will be represented by two vertices with the same name); order between keys will be represented with directed edges where between the same vertices we shall have edges in both directions. Let as denote this graph $G = (V, E)$, where $V$ is a multiset of vertices and $E$ set of edges.

---

[1] By abuse the notation, we shall write $\pi$ for the prefixes and the labels of the actions in both semantics, since they are essentially the same

To show exact correspondence between cause sets $K_F$ and $K_B$ we need to take the history part of the process $X$ involved in the execution of an action $\pi$. We can do it by taking all the paths in $G$ in which the target vertex will be key $i$ of the action $\pi$ and we shall obtain the graph $G(i) = (V(i), E(i))$. The multiset $V(i)$ contains all the keys which cause the action $\pi$ including $i$ and we can conclude that $K_F = V(i) \setminus \{i\}$. By removing all bidirectional edges from the graph $G(i)$ and replacing vertices that they connect with vertex renamed to $\tau_l$ when $l = 1, 2, \ldots n$, we shall obtain the graph $G' = (V', E')$. (Renaming is applied also on the other edges containing removed vertices. The operation of bidirectional edge contraction is precisely defined in Appendix A.3.) The set $V'$ differs from $V(i)$ in having $\tau_l$ vertices instead of the pairs of the vertices with the same name (originally belonging to silent moves in the framework). Hence, we can conclude that $K_B = V' \setminus (\{i\} \cup \tau_l)$.

We shall call the algorithm explained above 'Removing Keys from a Set', denoted as `Rem`. We shall write $\texttt{Rem}(K_F) = K_B$, meaning that $K_B$ can be obtained by applying algorithm `Rem` to $K_F$.

Before stating the theorem, we shall give a definition of the erasing function $\lambda$ and the function $\gamma$ that maps labels from the framework into labels from Boreale and Sangiorgi's semantics:

**Definition 22.** *The function $\gamma$ that maps label from the framework with a label from Boreale and Sangiorgi's semantics, is inductively defined as follows:*

$$\gamma((i, K, j) : \pi) = i : \gamma(\pi) \quad \text{when } \pi \neq \tau \qquad \gamma((i, *, *) : \tau) = \tau$$
$$\gamma(\overline{b}\langle va_\Delta\rangle) = \overline{b}\langle va\rangle \quad \text{when } \texttt{empty}(\Delta) = true \qquad \gamma(b(c)) = b(c)$$
$$\gamma(\overline{b}\langle va_\Delta\rangle) = \overline{b}a \quad \text{when } \texttt{empty}(\Delta) = false \qquad \gamma(\overline{b}a) = \overline{b}a$$

**Definition 23.** *The erasing function $\lambda$ that maps causal processes from Boreale and Sangiorgi's semantics to the $\pi$-calculus is inductively defined as follows:*

$$\lambda(A \mid A') = \lambda(A) \mid \lambda(A') \qquad \lambda(\texttt{K} :: A) = \lambda(A) \qquad \lambda(va(A)) = va(\lambda(A)) \qquad \lambda(P) = P$$

Now we have all necessary definitions to state the lemma about structural correspondence between two causal semantics.

**Lemma 5** (Structural correspondence). *Starting from initial $\pi$-calculus process P, where $P = A_1 = X_1$, we have:*

1. *if $P \xrightarrow[K_{B1}]{\zeta_1} A_2 \ldots A_n \xrightarrow[K_{Bn}]{\zeta_n} A_{n+1}$ then there exists a trace $P \xrightarrow[K_{F1}]{\mu_1}\!\!\!\twoheadrightarrow X_2 \ldots X_n \xrightarrow[K_{Fn}]{\mu_n}\!\!\!\twoheadrightarrow X_{n+1}$ and $K_{Fi}$, such that for all $i$, $\lambda(A_i) = \varphi(X_i)$, $\zeta_i = \gamma(\mu_i)$ and $\texttt{Rem}(K_{Fi}) = K_{Bi}$, for $i = 1, \ldots, n$.*

2. *if $P \xrightarrow[K_{F1}]{\mu_1}\!\!\!\twoheadrightarrow X_2 \ldots X_n \xrightarrow[K_{Fn}]{\mu_n}\!\!\!\twoheadrightarrow X_{n+1}$ then there exists a trace $P \xrightarrow[K_{B1}]{\zeta_1} A_2 \ldots A_n \xrightarrow[K_{Bn}]{\zeta_n} A_{n+1}$ where for all $i$, $\lambda(A_i) = \varphi(X_i)$, $\zeta_i = \gamma(\mu_i)$ and $\texttt{Rem}(K_{Fi}) = K_{Bi}$, for $i = 1, \ldots, n$.*

Considering the object dependence we have that the first action which extrudes a bound name will cause all future actions using that name in any position of the label. The main difference is that object dependence induced by input action in Boreale and Sangiorgi's semantics is subject dependence as well.

The next theorem demonstrates causal correspondence between causality in the framework when memory $\Delta$ is instantiated with $\Gamma_w$ and Boreale and Sangiorgi's late causal semantics.

**Theorem 2** (Causal correspondence). *The reflexive and transitive closure of the causality introduced in [6] coincides with the causality of the framework when $\Delta = \Gamma_w$.*

## 5 Conclusions

In a concurrent setting, causally-consistent reversibility relates causality and reversibility. Several works [10, 15, 6, 9, 8] have addressed causal semantics for π-calculus, differing on how object causality is modelled. Starting from this observation, we have devised a framework for reversible π-calculus which abstracts away from the underlying data structure used to record causes and consequences of an extrusion, and hence from the object causality. Depending on the underlying data structure, we can obtain different causal semantics. We have shown how three different semantics [10, 6, 9] can be derived, and we have proved causal correspondence with the semantics introduced in [6]. Our framework enjoys typical properties for reversible process algebra, such as loop lemma and causal consistence. As a future work we plan to prove causal correspondence with the semantics [10, 9] and to continue working towards a more parametric framework and to compare it with [25, 18]. Moreover it would be interesting to implement our framework in the psi-calculi framework [4], and to develop further the behavioural theory of our framework.

## Acknowledgments

We are grateful to the EXPRESS anonymous reviewers for theirs useful remarks and suggestions which led to substantial improvements.

## References

[1] L. Aceto (1994): *GSOS and Finite Labelled Transition Systems*. Theor. Comput. Sci. 131(1), pp. 181–195, doi:10.1016/0304-3975(94)90094-9.

[2] A. Avizienis, J-C Laprie, B. Randell & C. E. Landwehr (2004): *Basic Concepts and Taxonomy of Dependable and Secure Computing*. IEEE Trans. Dependable Sec. Comput. 1(1), pp. 11–33, doi:10.1109/TDSC.2004.2.

[3] G. Bacci, V. Danos & O. Kammar (2011): *On the Statistical Thermodynamics of Reversible Communicating Processes*. In: *CALCO 2011*, LNCS 6859, Springer, pp. 1–18, doi:10.1007/978-3-642-22944-2_1.

[4] J. Bengtson, M. Johansson, J. Parrow & B. Victor (2011): *Psi-calculi: a framework for mobile processes with nominal data and logic*. Logical Methods in Computer Science 7(1), doi:10.2168/LMCS-7(1:11)2011.

[5] C.H. Bennett (1973): *Logical Reversibility of Computation*. IBM Journal of Research and Development 17(6), doi:10.1147/rd.176.0525.

[6] M. Boreale & D. Sangiorgi (1998): *A Fully Abstract Semantics for Causality in the π-Calculus*. Acta Inf. 35(5), pp. 353–400, doi:10.1007/s002360050124.

[7] G. Boudol & I. Castellani (1988): *Permutation of transitions: An event structure semantics for CCS and SCCS*. In: *Linear Time, Branching Time and Partial Order in Logics and Models for Concurrency*, LNCS 354, Springer, pp. 411–427, doi:10.1007/BFb0013028.

[8] N. Busi & R. Gorrieri (1995): *A Petri Net Semantics for pi-Calculus*. In: *CONCUR Philadelphia, PA, USA, August 21-24, 1995, Proceedings*, pp. 145–159, doi:10.1007/3-540-60218-6_11.

[9] S. Crafa, D. Varacca & N. Yoshida (2012): *Event Structure Semantics of Parallel Extrusion in the Pi-Calculus*. In: *FOSSACS 2012*, LNCS 7213, Springer, pp. 225–239, doi:10.1007/978-3-642-28729-9_15.

[10] I. Cristescu, J. Krivine & D. Varacca (2013): *A Compositional Semantics for the Reversible π-Calculus*. In: *LICS 2013*, pp. 388–397, doi:10.1109/LICS.2013.45.

[11] I. D. Cristescu, J. Krivine & D. Varacca (2015): *Rigid Families for CCS and the π-calculus*. In: *ICTAC*, LNCS 9399, Springer, pp. 223–240, doi:10.1007/978-3-319-25150-9_14.

[12] V. Danos & J. Krivine (2004): *Reversible Communicating Systems*. In: *CONCUR 2004*, LNCS 3170, Springer, pp. 292–307, doi:10.1007/978-3-540-28644-8_19.

[13] V. Danos & J. Krivine (2005): *Transactions in RCCS*. In: *CONCUR 2005, San Francisco, CA, USA, August 23-26, 2005*, pp. 398–412, doi:10.1007/11539452_31.

[14] V. Danos & J. Krivine (2007): *Formal Molecular Biology Done in CCS-R*. Electr. Notes Theor. Comput. Sci. 180(3), pp. 31–49, doi:10.1016/j.entcs.2004.01.040.

[15] P. Degano & C. Priami (1999): *Non-Interleaving Semantics for Mobile Processes*. Theor. Comput. Sci. 216(1-2), pp. 237–270, doi:10.1016/S0304-3975(99)80003-6.

[16] E. Giachino, I. Lanese & C. A. Mezzina (2014): *Causal-Consistent Reversible Debugging*. In: *FASE 2014*, LNCS 8411, Springer, pp. 370–384, doi:10.1007/978-3-642-54804-8_26.

[17] J. Grattage (2005): *A Functional Quantum Programming Language*. In: *LICS*, IEEE Computer Society, Washington, DC, USA, pp. 249–258, doi:10.1109/LICS.2005.1.

[18] T. T. Hildebrandt, C. Johansen & H. Normann (2017): *A Stable Non-interleaving Early Operational Semantics for the Pi-Calculus*. In: *LATA*, LNCS 10168, pp. 51–63, doi:10.1007/978-3-319-53733-7_3.

[19] I. Lanese, M. Lienhardt, C. A. Mezzina, A. Schmitt & J. B. Stefani (2013): *Concurrent Flexible Reversibility*. In: *ESOP 2013*, pp. 370–390, doi:10.1007/978-3-642-37036-6_21.

[20] I. Lanese, C. A. Mezzina & J. B. Stefani (2016): *Reversibility in the higher-order $\pi$-calculus*. Theor. Comput. Sci. 625, pp. 25–84, doi:10.1016/j.tcs.2016.02.019.

[21] J-J Lévy (1976): *An algebraic interpretation of the $\lambda\beta K$-calculus; and an application of a labelled $\lambda$-calculus*. Theoretical Computer Science 2(1), pp. 97 – 114, doi:https://doi.org/10.1016/0304-3975(76)90009-8.

[22] D. Medic & C. A. Mezzina (2016): *Static VS Dynamic Reversibility in CCS*. In: *Reversible Computation RC 2016*, LNCS 9720, Springer, pp. 36–51, doi:10.1007/978-3-319-40578-0_3.

[23] D. Medic & C. A. Mezzina (2017): *Towards Parametric Causal Semantics in $\pi$-calculus*. In: *Joint Proceedings of the 18th Italian Conference on Theoretical Computer Science and the 32nd Italian Conference on Computational Logic, Naples, Italy, September 26-28.*, pp. 121–125.

[24] R. Milner (1980): *A Calculus of Communicating Systems*. LNCS 92, Springer, doi:10.1007/3-540-10235-3.

[25] R. Perera & J. Cheney (2017): *Proof-relevant $\pi$-calculus: a constructive account of concurrency and causality*. Mathematical Structures in Computer Science, pp. 1–37, doi:10.1017/S096012951700010X.

[26] I. Phillips, I. Ulidowski & S. Yuen (2013): *Modelling of Bonding with Processes and Events*. In: *Reversible Computation - RC 2013*, LNCS 7948, Springer, pp. 141–154, doi:10.1007/978-3-642-38986-3_12.

[27] I. C. C. Phillips & I. Ulidowski (2007): *Reversing algebraic process calculi*. J. Log. Algebr. Program. 73(1-2), pp. 70–96, doi:10.1016/j.jlap.2006.11.002.

[28] D. Sangiorgi & D. Walker (2001): *The Pi-Calculus - a theory of mobile processes*. Cambridge Uni. Press.

[29] M. V. Zelkowitz (1973): *Reversible Execution*. Commun. ACM 16(9), pp. 566–, doi:10.1145/362342.362360.

# A  Appendix

## A.1  Correspondence with the π-calculus

**(Lemma 1).** *If there is a transition $X \xrightarrow{\mu} Y$ then $\varphi(X) \xrightarrow{\varphi(\mu)}_\pi \varphi(Y)$.*

*Proof.* The proof is by induction on the derivation tree of the transition $X \xrightarrow{\mu} Y$. □

**(Lemma 2).** *If there is a transition $P \xrightarrow{\varphi(\mu)}_\pi Q$ then for all reachable $X$ such that $\varphi(X) = P$, there is a transition $X \xrightarrow{\mu} Y$ with $\varphi(Y) = Q$.*

*Proof.* The proof is by induction on the derivation tree of the transition $P \xrightarrow{\varphi(\mu)} Q$. □

## A.2  The main properties of the framework

In the following we shall give the proofs of our main theorems and additional necessary lemmas.

**Loop Lemma (Lemma 3).** *For every reachable process $X$ and forward transition $t : X \xrightarrow{\mu} Y$ there exists a backward transition $t' : Y \xleftarrow{\mu} X$, and conversely.*

*Proof.* The proof follows from the symmetry of the forward and the backward rules. □

**Lemma 6.** *Two derivation trees have the same conclusion if and only if they have the same premises.*

*Proof.* Let us consider two derivation trees with the same conclusions:

$$\frac{p_1}{X \xrightarrow{\mu} Y} \qquad \frac{p_2}{X \xrightarrow{\mu} Y}$$

If $p_1 = p_2$ then by the construction of the rules we can notice that both derivations will have the same conclusion. To prove the opposite direction we proceed with the induction on the derivation tree of the transition $X \xrightarrow{\mu} Y$. For each rule we need to show that only one premise is possible. We shall present only the interesting cases. (Rules IN2, OUT2, PAR, RES are possible just if the premise has exactly the same label on the transition as the conclusion and we have $p_1 = p_2$.)

- (CAUSE REF) Having a transition

$$\frac{p_q}{va_\Delta(X) \xrightarrow{(i,K',j):\pi} Y} \quad a \in sub(\pi) \quad q \in \{1,2\}$$

the only applicable rule is CAUSE REF with $Y = va_\Delta X'_{[K'/K_q]@i}$, and we have

$$\frac{X \xrightarrow{(i,K_q,j):\pi} X'}{va_\Delta(X) \xrightarrow{(i,K',j):\pi} va_\Delta X'_{[K'/K_q]@i}} \quad a \in sub(\pi) \quad q \in \{1,2\} \; \mathtt{Cause}(\Delta,K_q,K')$$

Now we want to show that $K_1 = K_2$. Depending on the data structure used for $\Delta$, we have different definitions for the predicate $\mathtt{Cause}(\Delta, K_q, K')$:

  - if $\Delta$ is a set $\Gamma$, then we have $K_q = K'$ or $\exists K' \in \Gamma \; K_q \leadsto_X K'$.
    * If $K_q = K'$, then we have $K_1 = K_2 = K'$.

* If $\exists K' \in \Gamma\ K_q \leadsto_X K'$ then by definition of instantiation relation and design of the communication rules, name $a$ substituted some variable $x$ in the actions identified with $K_q$. Since substitution is unique for a name, we have $K_1 = K_2$.
  – if $\Delta$ is an indexed set $\Gamma_w$, we have $K' = K_q \cup \{w\}$, i.e. $K_1 \cup \{w\} = K_2 \cup \{w\}$. We can distinguish two cases, depending if $w$ is a $*$ or not:
    * if $w = *$, we have $K_1 = K_2$
    * if $w = k$, where $k \neq *$, we need to prove that $K_1 \cup \{k\} = K_2 \cup \{k\}$. Since it is an operation on sets, we need to prove that the key $k$ belongs to both sets $K_1$ and $K_2$ or neither one of them. By design of the framework, all restrictions of the name $a$ are nested. Once name $a$ is extruded with the action $k$, every following restriction will have $k$ as an index of the set $\Gamma$ (for example, in the process $va_{\Gamma_{k_1}}(va_{\Gamma'_{k_2}}X)$, the options for $k_1$ and $k_2$ are: $k_1 = k_2$ or $k_1 = k \wedge k_2 = *$, for some $k$). Both transitions $K_1$ and $K_2$ are executed on the same component in the process $X$ and they need to pass the same restrictions on the name $a$. Since $w = k$, the only possibilities are $K_1 = K_2 = \{*\}$ or $K_1 = K_2 = \{*, k\}$.
  – if $\Delta$ is a set indexed with a set $\Gamma_\Omega$, then we have $K' = K_q \cup \Omega$, i.e. $K_1 \cup \Omega = K_2 \cup \Omega$. The proof is similar to the one above but we need to prove it for every element in a set $\Omega$.

* (OPEN) Having a transition

$$\frac{p_q}{va_\Delta(X) \xrightarrow{(i,K',j):\overline{b}\langle va_\Delta \rangle} Y} \quad q \in \{1,2\}$$

the only applicable rule is OPEN and we have $Y = va_{\Delta+i}X'_{[K'/K_q]@i}$ with a rule

$$\frac{X \xrightarrow{(i,K_q,j):\pi_q} X' \quad \pi_q = \overline{b}a \quad \pi_q = \overline{b}\langle va_{\Delta'}\rangle \quad \texttt{Update}(\Delta, K_q, K')}{va_\Delta(X) \xrightarrow{(i,K',j):\overline{b}\langle va_\Delta \rangle} va_{\Delta+i}X'_{[K'/K_q]@i}}$$

Now we want to show that $K_1 = K_2$ and $\pi_1 = \pi_2$. Depending on the data structure used for $\Delta$, we have different definitions for the condition $\texttt{Update}(\Delta, K_q, K')$:

  – if $\Delta$ is a set $\Gamma$, we have $K_1 = K_2 = K'$ and we just need to prove that $\pi_1 = \pi_2$. Let us suppose that $\pi_1 = \overline{b}a$; then this implies that there does not exist a context $C[\bullet]$ such that $X = C[va_{\Gamma'}X'']$, hence $\pi_2 = \overline{b}a$. Similarly for $\pi_1 = \overline{b}\langle va_{\Gamma'}\rangle$.
  – if $\Delta = \Gamma_\Omega$, the cause does not change with the rule OPEN and the proof is similar to the case above.
  – if $\Delta$ is an indexed set $\Gamma_w$, we have $K' = K_q \cup \{w\}$ and we need to prove that $K_1 = K_2$ and $\pi_1 = \pi_2$. Let us suppose that $\pi_1 = \overline{b}a$; this implies that there does not exist a context $C[\bullet]$ such that $X = C[va_{\Gamma'_w}X'']$, and we can conclude that $\pi_2 = \overline{b}a$ and $K_1 = K_2 = *$. Let us suppose that $\pi_1 = \overline{b}\langle va_{\Gamma'_w}\rangle$; then this implies that there exists a context $C[\bullet]$ such that $X = C[va_{\Gamma_w}X'']$, and we can conclude that $\pi_2 = \overline{b}\langle va_{\Gamma'_{w'}}\rangle$. For the cause sets $K_q$ we have two possibilities: if $w' = *$, we have $K_q = \{*\}$; if $w' = k$, we have $K_q = \{*, k\}$.

* (CLOSE): Having a transition

$$\frac{p_q}{X \mid Y \xrightarrow{(i,*,*):\tau} Z} \quad q \in \{1,2\}$$

If there exists $va_\Delta \in X$, we can apply only the rule CLOSE and get $Z = va_\Delta(X'_{\#i} \mid Y')$ with a rule:

$$\dfrac{X \xrightarrow{(i,K_q,j):\overline{b}\langle va_\Delta\rangle} X' \quad Y \xrightarrow{(i,K'_q,j'):b(x)} Y'}{X \mid Y \xrightarrow{(i,*,*):\tau} va_\Delta(X'_{\#i} \mid Y')} \quad K_q =_* j' \wedge K'_q =_* j$$

Now we want to show that $K_1 = K_2$. Depending on the data structure used for $\Delta$, we have different cases:

- if $\Delta$ is a set $\Gamma$, we have $K_q =_* j' \wedge K'_q =_* j$ and two cases:
  * if $b$ is a free name, then $K_q = K'_q = *$
  * if $b$ is not a free name, then there exists a context $X = C'[vb_\Gamma X'']$ and we can conclude $K_1 \neq *, K_2 \neq *$ and $K'_1 = *, K'_2 = *$. In order to satisfy the condition of the rule, we have $K_1 = K_2 = j'$.
- if $\Delta$ is an indexed set $\Gamma_w$, the action $\overline{b}\langle va_{\Gamma_w}\rangle$ can be caused through subject and object position in the label, and we need to reason on both names: $a$ and $b$. We shall divide cause sets $K_q$ into two subsets: $K_q = K_{q_a} \cup K_{q_b}$. Since the rule CLOSE is used, there exists a context $C[\bullet]$, such that $X = C[va_{\Gamma_w} X'']$ and we have two possibilities:
  * if there does not exist a context $X = C'[vb_{\Gamma'_{w'}} X''']$, we have $K_{q_b} = *$. For $K_{q_a}$ we have two cases: if $w = *$, we can conclude that $K_{q_a} = *$ and we have $K_1 = K_2$; if $w = k''$ for some $k'' \in \Gamma$, we can conclude that $K_{q_a} = \{*, k''\}$ and $K_1 = K_2$.
  * if there exists a context $X = C'[vb_{\Gamma'_{w'}} X''']$ then we have that $w' = k'''$ and $K_{q_b} = \{k''', *\}$. With the same reasoning as in the case above we have that $K_1 = K_2$.
- if $\Delta = \Gamma_\Omega$, we have two cases:
  * if $b$ is a free name, then $K_q = K'_q = *$
  * if $b$ is not a free name, there exists a context $X = C'[vb_{\Gamma'_{\Omega'}} X'']$ and we can conclude that $K_q = \Omega$ and $K'_q = *$.

- The proof for the rule COM is similar to the one for the rule CLOSE.
- <span style="color:red">The proof for the backward rules follows from the proof for the forward rules (all forward transition have a unique derivation) and Loop Lemma (Lemma 3).</span>

□

In a reversible process, we shall call the sequence of the actions executing on one component of the parallel composition a *branch*. By the structure of the reversible process (and π-calculus in general) branches can be nested. For example, in the process $X = \pi[1,*].(\pi'[2,*].(P_1 \mid P_2 \mid P_3) \mid Q)$, by observing the first branching, we have two branches: $\pi'[2,*].(P_1 \mid P_2 \mid P_3)$ and $Q$. A path in a reversible process, starting from the very first prefix and ending with the last one that could be executed on some component in the parallel composition, we shall call a *thread*. Using the same process $X$ as in example above, we have four threads: $T_1 = \pi[1,*].\pi'[2,*].P_1$, $T_2 = \pi[1,*].\pi'[2,*].P_2$, $T_3 = \pi[1,*].\pi'[2,*].P_3$ and $T_4 = \pi[1,*].Q$. We can notice that thread is defined on the process $X$, not depending on the branching point. The prefixes in a thread are always structurally dependent.

**Square Lemma (Lemma 4).** *If $t_1 : X \xrightarrow{\mu_1} Y$ and $t_2 : Y \xrightarrow{\mu_2} Z$ are two concurrent transitions, there exist $t'_2 : X \xrightarrow{\mu'_2} Y_1$ and $t'_1 : Y_1 \xrightarrow{\mu'_1} Z$ where $\mu_i =_\lambda \mu'_i$. We shall follow the standard notation and say that $t_2$ is a residual of $t'_2$ after $t_1$, denoted with $t_2 = t'_2/t_1$.*

*Proof.* The proof is by case analysis on the form of the transitions $t_1$ and $t_2$. We shall consider four cases on whether transitions $t_1$ and $t_2$ are synchronisations or not and show just the interesting one, when a reversible process is written in the form $X = \nu \tilde{a}_\Delta (X_1 \mid X_2)$ and a restriction of a name is involved. Before starting with the proof, let us make some comments:

**C1** Having a transition $t$ that involves just one branch, we can rewrite process $X$ as

$$X = \nu \widetilde{a_{n\Delta_n}} C_n[\ldots \nu \widetilde{a_{0\Delta_0}} C_0[X_1]]$$

where branches are represented with the contexts $C_0, \ldots C_n$ which do not contain restrictions on $a$ and $\nu \widetilde{a_{l\Delta_l}} = \nu \widetilde{a_{l1\Delta_{l1}}} \ldots \nu \widetilde{a_{ln\Delta_{ln}}}$. For the sake of simplicity we shall assume that the vector of names $\nu \widetilde{a_{l\Delta_l}}$ is a singleton and we can write:

$$X = \nu a_{\Delta_n} C_n[\ldots \nu a_{\Delta_0} C_0[X_1]]$$

When transition $t$ is not a synchronisation, we can distinguish two cases depending on the nature of the action $\pi$ in the transition $t$: if $\pi \neq \overline{b}\langle \nu a_\Delta \rangle$ for some name $b$, transition $t$ modifies just its own branch, not restrictions before it. Supposing that $t$ modifies branch $X_1$, we have

$$t : X \xrightarrow{\mu} \nu a_{\Delta_n} C_n[\ldots \nu a_{\Delta_0} C_0[X_1']]$$

if $\pi = \overline{b}\langle \nu a_\Delta \rangle$, transition $t$ modifies its branch and all the restrictions on $a$ and we have

$$t : X \xrightarrow{\mu} \nu a_{\Delta_n'} C_n[\ldots \nu a_{\Delta_0'} C_0[X_1']]$$

**C2** By having a synchronisation $t$ which involves just one branch in the context $C$, from transition $C[\widetilde{\nu a_\Delta} Y] \xrightarrow{\mu} C[\widetilde{\nu a_\Delta'} Y']$ we can conclude that context $C[\bullet]$ does not change, just the branch inside of it.

If we consider that $t$ involves two contexts $C_i[\bullet], C_j[\bullet]$ and that name $a$ is used in object position, rules CLOSE or CLOSE$^\bullet$ can be applied. We shall have the following transitions:

$$t : \nu a_{\Delta_n} C_n[\ldots \nu a_{\Delta_i} C_i[\ldots \nu a_{\Delta_j} C_j[\nu a_{\Delta_0} C_0[X_1]]]] \xrightarrow{\mu} \nu a_{\Delta_n} C_n[\ldots \nu a'_{\Delta_i'} C_i'[\ldots \nu a_{\Delta_j'} C_j'[\nu a_{\Delta_0} C_0[X_1]]]]$$

We can notice that synchronisation modifies involved contexts and restrictions on $a$ before them.

We proceed with case analysis on whether transitions $t_1$ and $t_2$ are synchronisations or not:

1. $t_1$ and $t_2$ are not synchronisations; we need to prove that by changing the order of the transitions we shall obtain the same process. We shall consider just the cases where transitions $t_1$ and $t_2$ modify not just their own branch, but also other contexts or restrictions. Let us assume that transition $t_1$ modifies branch $X_1$ and the performed action is $\pi_1 = \overline{b}\langle \nu a_\Delta \rangle$. Using note **C1** we have

$$t_1 : X \xrightarrow{\mu} \nu a_{\Delta_n'} C_n[\ldots \nu a_{\Delta_0'} C_0[X_1']]$$

Let $\pi_2$ be the action of the transition $t_2$. If $a \notin \pi_2$ then $t_2$ modifies just its own context (for example context $C_i$) and it is not prevented by the restrictions on $a$. Let us consider the case when $a \in \pi_2$. We continue the proof with the induction on the structure of the process.

- The base case of induction is to prove that if

$$va_\Delta(X_1 \mid X_2) \xrightarrow{\mu_1} va_{\Delta'_1}(X'_1 \mid X_2) \xrightarrow{\mu_2} va_{\Delta'}(X'_1 \mid X'_2)$$

then

$$va_\Delta(X_1 \mid X_2) \xrightarrow{\mu'_2} va_{\Delta'_2}(X_1 \mid X'_2) \xrightarrow{\mu'_1} va_{\Delta'}(X'_1 \mid X'_2)$$

Since $t_1$ has performed action $\pi_1 = \bar{b}\langle va_\Delta \rangle$, rules OPEN and OPEN$^\bullet$ can be used. We consider just the interesting cases when on $t_1$ is applied rule OPEN and continue with the case analysis on the rules that can be applied on $t_2$ such that $a \in \pi_2$.

– Rule OPEN applied on $t_2$. We have

$$va_\Delta(X_1 \mid X_2) \xrightarrow{(i_1,K_1,j_1):\bar{b}\langle va_\Delta\rangle} va_{\Delta+i_1}(X'_1 \mid X_2) \xrightarrow{(i_2,K_2,j_2):\bar{c}\langle va_{\Delta+i_1}\rangle} va_{\Delta+i_1+i_2}(X'_1 \mid X'_2)$$

Since the transition $t_1$ and $t_2$ are concurrent (i.e. there is no causal relation between them) we have that $i_1 \notin K_2$ and $i_1 \notin K$, where $X_2 \xrightarrow{(i_2,K,j_2):\bar{c}\langle va_{\Delta'}\rangle} X'_2$ (by definition of Update$(\cdot)$, having $i_1 \in K$ will lead to $i_1 \in K_2$, which is in contradiction with the fact that transitions are concurrent). Now we can safely commute transitions and we have:

$$va_\Delta(X_1 \mid X_2) \xrightarrow{(i_2,K_2,j_2):\bar{c}\langle va_\Delta\rangle} va_{\Delta+i_2}(X_1 \mid X'_2) \xrightarrow{(i_1,K_1,j_1):\bar{b}\langle va_{\Delta+i_2}\rangle} va_{\Delta+i_1+i_2}(X'_1 \mid X'_2)$$

as desired. The labels of transitions $t_1$ and $t'_1$ are not equal; they are equivalent, i.e. $\mu_1 =_\lambda \mu'_1$.

– Rule OPEN$^\bullet$ applied on $t_2$. We have

$$va_{\Delta+i_2}(X_1 \mid X_2) \xrightarrow{(i_1,K_1,j_1):\bar{b}\langle va_{\Delta+i_2}\rangle} va_{\Delta+i_1+i_2}(X'_1 \mid X_2) \rightsquigarrow^{(i_2,K_2,j_2):\bar{c}\langle va_{\Delta+i_1}\rangle} va_{\Delta+i_1}(X'_1 \mid X'_2)$$

Transition $t_2$ cannot be caused by transition $t_1$ since transition $t_1^\bullet$ (the forward one) happened before $t_1$. We can safely commute transitions and get:

$$va_{\Delta+i_2}(X_1 \mid X_2) \rightsquigarrow^{(i_2,K_2,j_2):\bar{c}\langle va_\Delta\rangle} va_\Delta(X_1 \mid X'_2) \xrightarrow{(i_1,K_1,j_1):\bar{b}\langle va_\Delta\rangle} va_{\Delta+i_1}(X'_1 \mid X'_2)$$

– Rule CAUSE REF applied on $t_2$. We have

$$va_\Delta(X_1 \mid X_2) \xrightarrow{(i_1,K_1,j_1):\bar{b}\langle va_\Delta\rangle} va_{\Delta+i_1}(X'_1 \mid X_2) \xrightarrow{(i_2,K_2,j_2):\pi} va_{\Delta+i_1}(X'_1 \mid X'_2)$$

with $X_2 \xrightarrow{(i_2,K,j_2):\pi} X'_2$ and $a \in sub(\alpha)$. Since the transitions $t_1$ and $t_2$ are concurrent, we have that $i_1 \notin K_2$ and $i_1 \notin K$. Now we can safely commute transitions and we have:

$$va_\Delta(X_1 \mid X_2) \xrightarrow{(i_2,K_2,j_2):\pi} va_\Delta(X_1 \mid X'_2) \xrightarrow{(i_1,K_1,j_1):\bar{b}\langle va_\Delta\rangle} va_{\Delta+i_1}(X'_1 \mid X'_2)$$

- In the inductive case we want to show that if $X \xrightarrow{\mu_1} Y \xrightarrow{\mu_2} Z$ and $X \xrightarrow{\mu'_2} Y_1 \xrightarrow{\mu'_1} Z$, then the following holds

$$va_\Delta X \xrightarrow{\mu_1} va_{\Delta'_1} Y \xrightarrow{\mu_2} va_{\Delta'} Z \quad \text{and} \quad va_\Delta X \xrightarrow{\mu'_2} va_{\Delta'_2} Y_1 \xrightarrow{\mu'_1} va_{\Delta'} Z$$

$$Z_i \mid X \xrightarrow{\mu_1} Z_i \mid Y \xrightarrow{\mu_2} Z_i \mid Z \quad \text{and} \quad Z_i \mid X \xrightarrow{\mu'_2} Z_i \mid Y_1 \xrightarrow{\mu'_1} Z_i \mid Z$$

Both subcases are straightforward.

2. Let consider that $t_2$ is a synchronisation and $t_1$ is not. We shall consider the case when $t_1$ is performing an action $\bar{b}\langle va_\Delta \rangle$ and it modifies the branch $X_1$ of the process

$$X = va_{\Delta_n} C_n[\ldots va_{\Delta_i} C_i[\ldots va_{\Delta_j} C_j[va_{\Delta_0} C_0[X_1]]]]$$

In this case, the applied rules could be OPEN or OPEN$^\bullet$. If $t_2$ is a synchronisation which does not involve the name $a$, then by note **C2**, we can conclude that it is trivial. We shall consider the case when $t_2$ uses name $a$ in object position and in this case, rules CLOSE or CLOSE$^\bullet$ are used. The contexts involved are $C_i[\bullet]$, $C_j[\bullet]$. Now we can proceed with an induction on the structure of the process.

- In the base case we shall have $X = va_{\Delta_n} C_n[\ldots va_{\Delta_i} C_i[\ldots va_{\Delta_j} C_j[va_{\Delta_0} C_0[X_1]]]]$ and for the sake of simplicity, we can rewrite it as $va_{\Delta_i} C_i[\ldots va_{\Delta_j} C_j[X_1]]$. Now we combine the rules applied on transition $t_1$ with the one applied on $t_2$ and we have:
  - Rule OPEN applied on $t_1$ and rule CLOSE on the $t_2$. We have:

  $$X = va_{\Delta_i} C_i[\ldots va_{\Delta_j} C_j[X_1]] \xrightarrow{(i_1,K_1,j_1):\bar{b}\langle va_{\Delta_i}\rangle} va_{\Delta_i+i_1} C_i[\ldots va_{\Delta_j+i_1} C_j[X_1']]$$
  $$\xrightarrow{(i_2,*,*):\tau} va_{\Delta_i+i_1} va_{\Delta'+i_1} C_i'[\ldots va_{\Delta_j'+i_1} C_j'[X_1']] = Z$$

  where $va_{\Delta+i_1} \in C_i[\ldots va_{\Delta_j+i_1} C_j[X_1']]$. If we commute transitions, we shall have:

  $$X = va_{\Delta_i} C_i[\ldots va_{\Delta_j} C_j[X_1]] \xrightarrow{(i_2,*,*):\tau} va_{\Delta_i} va_{\Delta'} C_i'[\ldots va_{\Delta_j'} C_j'[X_1]]$$
  $$\xrightarrow{(i_1,K_1,j_1):\bar{b}\langle va_{\Delta_i}\rangle} va_{\Delta_i+i_1} va_{\Delta'+i_1} C_i'[\ldots va_{\Delta_j'+i_1} C_j'[X_1']] = Z$$

  where $va_\Delta \in C_i[\ldots va_{\Delta_j} C_j[X_1]]$. We have $Z = Z$, as desired.
  - Rule OPEN$^\bullet$ applied on $t_1$ and rule CLOSE on the $t_2$. We have:

  $$X = va_{\Delta_i+i_1} C_i[\ldots va_{\Delta_j+i_1} C_j[X_1]] \stackrel{(i_1,K_1,j_1):\bar{b}\langle va_{\Delta_i}\rangle}{\rightsquigarrow} va_{\Delta_i} C_i[\ldots va_{\Delta_j} C_j[X_1']]$$
  $$\xrightarrow{(i_2,*,*):\tau} va_{\Delta_i} va_{\Delta'} C_i'[\ldots va_{\Delta_j'} C_j'[X_1']] = Z$$

  where $va_\Delta \in C_i[\ldots va_{\Delta_j} C_j[X_1']]$. By changing the order of the transitions, we have:

  $$X = va_{\Delta_i+i_1} C_i[\ldots va_{\Delta_j+i_1} C_j[X_1]] \xrightarrow{(i_2,*,*):\tau} va_{\Delta_i+i_1} va_{\Delta'+i_1} C_i'[\ldots va_{\Delta_j'+i_1} C_j'[X_1]]$$
  $$\stackrel{(i_1,K_1,j_1):\bar{b}\langle va_{\Delta_i}\rangle}{\rightsquigarrow} va_{\Delta_i} va_{\Delta'} C_i'[\ldots va_{\Delta_j'} C_j'[X_1']] = Z$$

  where $va_{\Delta+i_1} \in C_i[\ldots va_{\Delta_j+i_1} C_j[X_1]]$.
  - Similarly for the rules OPEN and OPEN$^\bullet$ combined with rule CLOSE$^\bullet$.
- The inductive case is trivial since only $t_1$ modifies processes in the context, not the context by itself. (Considering the synchronisation $X_1 \xrightarrow{\mu} X_1'$, we have $C[va_\Delta(X_1 \mid X_2)] \xrightarrow{\mu} C[va_\Delta(X_1' \mid X_2)]$ where we can notice that the transition modifies just $X_1$.)

3. The case when $t_1$ is a synchronisation and $t_2$ is not, is similar to the one above.

4. $t_1$ and $t_2$ are synchronisations.

   Let suppose that $t_1$ is a synchronisation between branch with output in $X_1$ and branch with input in the context $C_j$, while $t_2$ is a synchronisation between branches with input in $C_i$ and output in $C_k$. Since transitions will modify contexts just up to $\nu a_{\Delta_i}$, we can reason on process $X$ written as:

   $$X = \nu a_{\Delta_i} C_i[\ldots \nu a_{\Delta_j} C_j[\ldots \nu a_{\Delta_k} C_k[X_1]]]$$

   We continue with the case analysis depending whether name $a$ is in the subject or in the object position in the transitions $t_1$ and $t_2$.

   - name $a$ is in the object position in both transitions; in this case, rules CLOSE and CLOSE$^\bullet$ can be used. Let us consider the case when rule CLOSE is applied on $t_1$ and on $t_2$; the rest of the cases are similar. We have

     $$\nu a_{\Delta_i} C_i[\ldots \nu a_{\Delta_j} C_j[\ldots \nu a_{\Delta_k} C_k[X_1]]] \xrightarrow{(i_1,*,*):\tau}$$
     $$\nu a_{\Delta_i} C_i[\ldots \nu a_{\Delta_j} \nu a_{\Delta'} C'_j[\ldots \nu a_{\Delta'_k} C_k[X'_1]]] \xrightarrow{(i_2,*,*):\tau} \nu a_{\Delta_i} \nu a_{\Delta''_l} C'_i[\ldots \nu a_{\Delta'_j} \nu a_{\Delta''} C'_j[\ldots \nu a_{\Delta'_k} C'_k[X'_1]]]$$

     where $\nu a_\Delta \in C_j[\ldots \nu a_{\Delta_k} C_k[X_1]]$ and $\nu a_{\Delta'_l} \in C_i[\ldots \nu a_{\Delta_j} \nu a_{\Delta'} C'_j[\ldots \nu a_{\Delta'_k} C_k[X'_1]]]$. The action $i_2$ cannot be caused by the action $i_1$ because the transitions are concurrent.
     By changing the order of the transitions we have that derivation of $t'_2$ is:

     $$\nu a_{\Delta_i} C_i[\ldots \nu a_{\Delta_j} C_j[\ldots \nu a_{\Delta_k} C_k[X_1]]] \xrightarrow{(i_2,*,*):\tau} \nu a_{\Delta_i} \nu a_{\Delta''_l} C'_i[\ldots \nu a_{\Delta'_j} C_j[\ldots \nu a_{\Delta'_k} C'_k[X_1]]]$$

     where $\nu a_{\Delta_l} \in C_i[\ldots \nu a_{\Delta_j} C_j[\ldots \nu a_{\Delta_k} C_k[X_1]]]$. The derivation for the transition $t'_1$ is:

     $$\nu a_{\Delta_i} \nu a_{\Delta''_l} C'_i[\ldots \nu a_{\Delta'_j} C_j[\ldots \nu a_{\Delta'_k} C'_k[X_1]]] \xrightarrow{(i_1,*,*):\tau} \nu a_{\Delta_i} \nu a_{\Delta''_l} C'_i[\ldots \nu a_{\Delta'_j} \nu a_{\Delta''} C'_j[\ldots \nu a_{\Delta''_k} C'_k[X'_1]]]$$

     where $\nu a_{\Delta'} \in C_j[\ldots \nu a_{\Delta'_k} C'_k[X_1]]$.

   - name $a$ is in the object position in transition $t_1$ and in subject in $t_2$. In this case, rules CLOSE and CLOSE$^\bullet$ can be applied on $t_1$ and COM and COM$^\bullet$ on $t_2$. Let us consider the case when CLOSE is applied on $t_1$ and COM on $t_2$; the rest of the cases are similar. We have

     $$\nu a_{\Delta_i} C_i[\ldots \nu a_{\Delta_j} C_j[\ldots \nu a_{\Delta_k} C_k[X_1]]] \xrightarrow{(i_1,*,*):\tau} \nu a_{\Delta_i} C_i[\ldots \nu a_{\Delta_j} \nu a_{\Delta'} C'_j[\ldots \nu a_{\Delta'_k} C_k[X'_1]]]$$

     where $\nu a_\Delta \in C_j[\ldots \nu a_{\Delta_k} C_k[X_1]]$. By executing the rule COM we are changing just contexts $C_i$ and $C_k$, not the restrictions.

     $$\nu a_{\Delta_i} C_i[\ldots \nu a_{\Delta_j} \nu a_{\Delta'} C'_j[\ldots \nu a_{\Delta'_k} C_k[X'_1]]] \xrightarrow{(i_2,*,*):\tau} \nu a_{\Delta_i} C'_i[\ldots \nu a_{\Delta_j} \nu a_{\Delta'} C'_j[\ldots \nu a_{\Delta'_k} C'_k[X'_1]]]$$

     The cause of the actions on the channel $a$ of the transition $t_2$ cannot choose as a cause $i_1$ since the transitions are concurrent. Hence, we can commute transitions and get:

     $$\nu a_{\Delta_i} C_i[\ldots \nu a_{\Delta_j} C_j[\ldots \nu a_{\Delta_k} C_k[X_1]]] \xrightarrow{(i_2,*,*):\tau}$$
     $$\nu a_{\Delta_i} C'_i[\ldots \nu a_{\Delta_j} C_j[\ldots \nu a_{\Delta_k} C'_k[X_1]]] \xrightarrow{(i_1,*,*):\tau} \nu a_{\Delta_i} C'_i[\ldots \nu a_{\Delta_j} \nu a_{\Delta'} C'_j[\ldots \nu a_{\Delta'_k} C'_k[X'_1]]]$$

     where $\nu a_\Delta \in C_j[\ldots \nu a_{\Delta_k} C'_k[X_1]]]$ since transition $t_2$ does not change restrictions.

□

Two transitions are *coinitial* if they have the same source, *cofinal* if they have the same target, and *composable* if the target of one is the source of the other one. A sequence of pairwise composable transitions is called a *trace*. We denote with $\varepsilon$ the *empty* trace, and with $t_1;t_2$ the composition of two composable traces $t_1$ and $t_2$. In the main theorem we prove that reversibility in our framework is causally consistent.

**Definition [Equivalence up-to permutation] (Definition 21).** *Equivalence up-to permutation, $\sim$, is the least equivalence relation on the traces, satisfying:*

$$t_1;t_2 \sim (t_2/t_1);(t_1/t_2) \qquad t;t^\bullet \sim \varepsilon$$

**Definition 24.** *Two transitions $t_1$ and $t_2$ are* prefix equivalent, *written $t_1 =_p t_2$ if they add or remove the same past element from the history of a process.*

**Example 7.** *Having the process $\overline{a}^*b^*.P_1 \mid \overline{a}^*b^*.P_2$, we have two transitions*

$$t_1 : \overline{a}^*b^*.P_1 \mid \overline{a}^*b^*.P_2 \xrightarrow{(i,*,*):\overline{a}b} \overline{a}^*b^*[i,*].P_1 \mid \overline{a}^*b^*.P_2$$

$$t_2 : \overline{a}^*b^*.P_1 \mid \overline{a}^*b^*.P_2 \xrightarrow{(i,*,*):\overline{a}b} \overline{a}^*b^*.P_1 \mid \overline{a}^*b^*[i,*].P_2$$

Transitions $t_1$ and $t_2$ are not the same, but they are prefix equivalent because they add the same past element into the history. The LTS ensures that keys are unique in the trace.

**Lemma 7.** *If transitions $t_1$ and $t_2$ are prefix equivalent, coinitial and they are on the same branch of a process, then $t_1 = t_2$.*

*Proof.* The proof follows from the fact that keys are unique and that transitions are coinitial. □

**Lemma 8** (Parabolic traces). *Let s be a trace. Then there exist a backward-only trace r and a forward-only trace $r'$ such that $s \sim r;r'$.*

*Proof.* The proof is by induction on the length of *s* and the distance between the very first transition in *s* and the pair of transitions contradicting the statement of the lemma. Let suppose that this is a pair of transitions $t_1, t_2$; then we have:

$$t_1 : X \xrightarrow{(i_1,k_1,j_1):\alpha_1} Y \qquad t_2 : Y \rightsquigarrow^{(i_2,k_2,j_2):\alpha_2} Z$$

We have two cases depending if the keys of $t_1$ and $t_2$ are the same or not.

- if $i_1 = i_2$, by the fact that keys are unique in a reversible process, the only possible option to execute transition $t_2$ with the same key as transition $t_1$ is if $t_2 = t_1^\bullet$ and we have $X = Z$. Hence, we can eliminate transitions $t_1$ and $t_2$ and decrease the length of *s*.
- if $i_1 \neq i_2$, then we have two possibilities:
  - $t_1$ and $t_2$ are concurrent; then we can apply Lemma 4 and swap them. In this way we decrease the distance between the very first transition in *s* and the pair of transitions contradicting the statement of the lemma.
  - $t_1$ and $t_2$ are causally dependent; this case is impossible. They cannot be structural or object dependent since transition $t_2$ is backward one and $t_1$ and $t_2$ are consecutive.

□

**Lemma 9.** *Let us denote with $s_1$ and $s_2$ two coinitial and cofinal traces, where $s_2$ is forward only. Then there exists a forward-only trace $s_1'$, shorter or equal to $s_1$, such that $s_1 \sim s_1'$.*

*Proof.* The proof is by induction on the length of $s_1$. If $s_1$ is forward-only then $s_1' = s_1$. If not, by Lemma 8, we can assume that $s_1$ is parabolic, and write it as $s_1 = u;t_1;t_2;v$ where $t_1,t_2$ is the only pair of consecutive transitions in the opposite direction; $u;t_1$ is backward-only and $t_2;v$ is forward-only. Since traces $s_1$ and $s_2$ are coinitial and cofinal and $s_2$ is a forward-only, we can notice that the history element which transition $t_1$ takes out of the history, some transition in $t_2;v$ needs to put back, otherwise, the difference will stay visible. Let us denote with $t'$ the first such transition. To preserve the same target in the end of the traces $s_1$ and $s_2$, we have that $t'$ is exact inverse of the transition $t_1$ i.e. $t' = t_1^{\bullet}$. We can rewrite $s_1$ as $u;t_1;t_2;v_1;t';v_2$.

We proceed by showing that $t_1$ is concurrent with all transitions up to $t'$. Suppose that there exists some $t''$ between $t_1$ and $t'$ such that $t_1$ and $t''$ are causal. Depending of the type of cause we can distinguish two cases:

- if $t_1$ and $t''$ are structural causal then we have a contradiction with the hypothesis that $t'$ is the first transition that will put back the history element that $t_1$ deletes.

- if $t_1$ and $t''$ are object causal then we have two possibilities:
    - $k_1 = i''$. This cannot happen since $t''$ is after $t_1$.
    - or $k'' = i_1$. This cannot happen since $t_1$ is backward.

We can conclude that transition $t_1$ is concurrent with all transitions between $t_1$ and $t'$. By Lemma 4 we can swap $t_1$ with each transitions up to $t'$ and we have $s_1 \sim u;t_2;v_1;t_1;t';v_2$. By definition of $\sim$ we have $s_1 \sim u;t_2;v_1;v_2$ (since $t_1^{\bullet} = t'$, we can erase the transitions if $t_1;t' \sim \varepsilon$). In this way, the length of $s_1$ decreases and we can apply the inductive hypothesis. $\square$

**Main Theorem (Theorem 1).** *Two traces are coinitial and cofinal if and only if they are equivalent up-to permutation.*

*Proof.* Let us denote two traces with $s_1$ and $s_2$. If $s_1 \sim s_2$ then from the definition of $\sim$ we can conclude that they are coinitial and cofinal. Let us suppose that $s_1$ and $s_2$ are coinitial and cofinal. From the Lemma 8 we can suppose that they are parabolic. We shall reason by induction on the lengths of $s_1, s_2$ and on the depth of the very first disagreement between them. We shall denote it with the pair $t_1, t_2$. Then we can write the traces $s_1$ and $s_2$ as

$$s_1 = u_1;t_1;v_1 \qquad s_2 = u_2;t_2;v_2$$

where $u_1 \sim u_2$. Depending on whether $t_1$ and $t_2$ are forward or not, we have the following cases:

- $t_1$ is forward and $t_2$ is backward. Since $s_1$ is parabolic we have that $u_1$ is backward-only and $v_1$ is forward-only. The traces $t_1;v_1$ and $t_2;v_2$ are coinitial and cofinal where $t_1;v_1$ is forward only. By Lemma 9, there exists $s_2'$ shorter or equal to $t_2;v_2$ such that $s_2' \sim t_2;v_2$. If it is equal then $t_2$ needs to be forward (since $s_2'$ is forward-only) and this is in contradiction with the hypothesis. If it is shorter then we proceed by induction on $u_2;s_2'$.

- $t_1$ and $t_2$ are forward. Then $t_1;v_1$ and $t_2;v_2$ are coinitial, cofinal and forward-only. We have two cases depending on whether $t_1$ and $t_2$ are concurrent or not.
    - if $t_1$ and $t_2$ are concurrent then whatever $t_1$ puts in the history, $v_2$ needs to do the same. Let $t_1'$ be the first such transition, then $t_1' \in v_2$ and $t_1' =_p t_1$. Now we can rewrite $t_2;v_2$ as $t_2;v_2';t_1';v_2''$ and show that $t_1'$ is concurrent with all transitions in $v_2'$:

* $t'_1$ is the first transition on the same thread as $t_1$. Hence, it is not structural causal with any transition in $t_2; v'_2$
* since $t_1$ is coinitial with $t_2; v'_2$ and $t_2; v'_2$ are forward-only, transition $t_1$ cannot have as contextual cause any transition from $t_2; v'_2$.

From Lemma 4 we have:

$$t_2; v_2 = t_2; v'_2; t'_1; v''_2 \sim t'_1; t_2; v'_2; v''_2.$$

Since $t'_1 =_p t_1$, they are on the same thread and they are coinitial, from Lemma 7 we have that $t'_1 = t_1$. Without changing the length of $s_1$ and $s_2$ we obtain the first disagreement pair later and we can rely on the inductive hypothesis.

- if $t_1$ and $t_2$ are causally related, then we need to check both types of causality:
  * structural causality: if $t_1 =_p t_2$, then by the Lemma 7, we have $t_1 = t_2$ which is a contradiction with the hypothesis. If $t_1 \neq_p t_2$ then the traces $t_1; v_1$ and $t_2; v_2$ are not cofinal, and this is not the case. Hence, $t_1$ and $t_2$ are not structural causal.
  * object causality: since $t_1$ and $t_2$ are both forward and coinitial, there cannot be object causality between them.

• The proof is similar if both $t_1$ and $t_2$ are backward.

□

## A.3 Correspondence with Boreale and Sangiorgi's semantics

In this section we give the definition of Boreale and Sangiorgi's late semantics and adapt the definition of object causality to it. After that, we show causal correspondence between Boreale and Sangiorgi's late semantics and the framework when memory $\Delta$ is instantiated with $\Gamma_w$.

**Boreale and Sangiorgi's late semantics.** The authors distinguish between two types of causality: *subject* and *object*. To capture the first one, they introduce a causal term $\mathtt{K} :: A$, defined on the top of the $\pi$-calculus as follows:

$$\text{(Causal process) } A, B ::= P \mid \mathtt{K} :: A \mid A \mid B \mid va(A)$$

where $\mathtt{K}$ is a set of causes recording the fact that every action performed by $A$ depends on $\mathtt{K}$; $P$ is classical $\pi$-calculus process. We denote by $Cau(A)$ the set of causes which appear in the causal process $A$. The main difference between the rules for the late semantics presented in Figure 5 and the original ones in [6] is that the name is substituted in the communication rules and not in the input rule. We use notation $A \xrightarrow{k:\eta}_{\mathtt{K}} A'$ instead of $A \xrightarrow{\eta}_{\mathtt{K};k} A'$, where $\eta = \bar{b}a \mid b(x) \mid \bar{b}\langle va\rangle \mid \tau$. When information about the type of the action and the cause are irrelevant, we write $A \xrightarrow{\zeta}_{\mathtt{K}} A'$, where $\zeta = k : \eta$. For the simplicity of the proofs, we divided the original rule COM from [6] into two rules: BS-COM and BS-CLOSE, where notation $A[k \rightsquigarrow \mathtt{K}]$ indicates the fact that cause $k$ needs to be replaced with a set $\mathtt{K}$.

To prove a causal correspondence, first we need to prove the structural correspondence. To do so, we observe traces in both semantics. Labels in the framework will bring additional information about the multiset of the structural causes of the executed action ($K_F$) and a trace in the framework will have the following form: $X_1 \xrightarrow{\mu_1}_{K_{F1}} X_2 \ldots X_{n-1} \xrightarrow{\mu_n}_{K_{Fn}} X_n$. The main difference is in the silent actions. In the framework, silent actions are identified with unique keys, while in Boreale and Sangiorgi's semantics,

$$(\text{BS-Out}) \ \overline{b}a.A \xrightarrow[\emptyset]{k:\overline{b}a} k :: A \qquad (\text{BS-In}) \ b(x).A \xrightarrow[\emptyset]{k:b(x)} k :: A \qquad (\text{BS-Cau}) \ \frac{A \xrightarrow[\text{K}]{k:\eta} A'}{\text{K}' :: A \xrightarrow[\text{K,K}']{k:\eta} \text{K}' :: A'}$$

$$(\text{BS-Com}) \ \frac{A_1 \xrightarrow[\text{K}_1]{k:\overline{b}a} A_1' \quad A_2 \xrightarrow[\text{K}_2]{k:b(x)} A_2' \quad k \notin Cau(A_1, A_2)}{A_1 \mid A_2 \xrightarrow{\tau} A_1'[k \rightsquigarrow \text{K}_2] \mid A_2'[k \rightsquigarrow \text{K}_1]} \qquad (\text{BS-Par}) \ \frac{A_1 \xrightarrow[\text{K}]{k:\eta} A_1' \quad k \notin Cau(A_2)}{A_1 \mid A_2 \xrightarrow[\text{K}]{k:\eta} A_1' \mid A_2}$$

$$(\text{BS-Open}) \ \frac{A \xrightarrow[\text{K}]{k:\overline{b}a} A'}{\nu a\, A \xrightarrow[\text{K}]{k:\overline{b}\langle \nu a\rangle} A'} \qquad (\text{BS-Res}) \ \frac{A \xrightarrow[\text{K}]{k:\eta} A' \quad a \notin \eta}{\nu a\, A \xrightarrow[\text{K}]{k:\eta} \nu a\, A'}$$

$$(\text{BS-Close}) \ \frac{A_1 \xrightarrow[\text{K}_1]{k:\overline{b}\langle \nu a\rangle} A_1' \quad A_2 \xrightarrow[\text{K}_2]{k:b(x)} A_2' \quad k \notin Cau(A_1, A_2)}{A_1 \mid A_2 \xrightarrow{\tau} \nu a(A_1'[k \rightsquigarrow \text{K}_2] \mid A_2'[k \rightsquigarrow \text{K}_1])}$$

Figure 5: Rules for Boreale and Sangiorgi's late semantics

they just merge the cause sets of the actions participating in the communication. Therefore, we need to provide a connection between sets of structural causes in these two semantics.

The idea is to represent the history of a reversible process by a directed graph and to use an edge contraction operation to obtain the cause set K of a causal process $A$. We can see causes $k$ in Boreale and Sangiorgi's semantics as identifiers of the visible actions, and we shall denote them with $i$. Before explaining the algorithm, we shall give a definition of a function $\gamma$ that maps labels from the framework into labels from Boreale and Sangiorgi's semantics:

**Definition 25** (Mapping function). *The function $\gamma$ that maps a label from the framework into a label from Boreale and Sangiorgi's semantics, is inductively defined as follows:*

$$\gamma((i, K, j) : \pi) = i : \gamma(\pi) \quad \text{when } \pi \neq \tau \qquad \gamma((i, *, *) : \tau) = \tau$$
$$\gamma(\overline{b}\langle \nu a_\Delta\rangle) = \overline{b}\langle \nu a\rangle \quad \text{when } \texttt{empty}(\Delta) = true \qquad \gamma(b(c)) = b(c)$$
$$\gamma(\overline{b}\langle \nu a_\Delta\rangle) = \overline{b}a \quad \text{when } \texttt{empty}(\Delta) = false \qquad \gamma(\overline{b}a) = \overline{b}a$$

Let us suppose that we have two transitions $t$ and $t'$, where $t : X \xrightarrow[K_F]{(i,K,j):\pi} X'$ and $t' : A \xrightarrow[K_B]{i:\eta} A'$, and where $\gamma((i, K, j) : \pi) = i : \eta$ and a continuation in the reversible process $X$ is the same one as in a causal process $A$ and it is $P$. We write $K_B$ instead of cause set K to be clearer about which cause set belongs to which semantics. We can represent the history of the process $X$ (all executed actions in the $X$) with a directed graph, in the following way: keys of executed actions will be represented as vertices of a graph (actions which are part of a communication and have the same key, will be represented by two vertices with the same name); order between keys will be represented by directed edges where between the same vertices we shall have edges in both directions. To make this precise:

**Definition 26.** *The history of the reversible process $X$ (all executed actions in $X$) can be represented as a*

*directed graph $G = (V,E)$, in the following way:*

- $\forall\, \alpha[i,K] \in X \implies i \in V$
- $\forall\, \alpha[i,K].\alpha'[i',K'] \in X \implies (i,i') \in E'$
- $E = E' \cup \{(i,i') \mid \text{when } i = i'\} \cup \{(i',i) \mid \text{when } i' = i\}$

*where V is a multiset of vertices and E a set of directed edges.*

A *path s* is a sequence of vertices and directed edges from $G$, where the first vertex is a *source* vertex of a path and the last one is a *target* vertex of a path. A *subgraph*, $G(h) = (V(h), E(h))$, is obtained from the original graph $G$ by taking all the paths $s_n$ such that vertex $h$ is the target vertex of the paths $s_n$. Considering transitions $t$ and $t'$ with their key $i$ defined before and observing subgraph $G(i) = (V(i), E(i))$, we have that $G(i)$ will represent the order between the causes of the transition $t$ in the framework. Hence, vertex set $V' \setminus i$ will correspond to the multiset of structural causes, $K_F$ of the transition $t$. Since bidirectional edges represent dependency flow between vertices with the same name, we can remove them and join two vertices into one, renamed to $\tau$. This operation is known as edge contraction and we shall here adapt it to bidirectional edges.

**Definition 27.** *Bidirectional edge contraction is an operation defined on the directed graph $G = (V,E)$, as follows:*

- $E' = E \setminus ((i_1,i_2) \cup (i_2,i_1))$ *when $i_1 = i_2$*
- $V' = (V \setminus \{i_1,i_2\}) \cup \{\tau\}$
- $\forall (i,i_l), (i_l,i) \in E$ *where $l \in \{i_1,i_2\}$, we have that $(i,\tau), (\tau,i) \in E'$,*

*where $G' = (V', E')$ is the obtained subgraph.*

By applying bidirectional edge contraction on every bidirectional edge of a graph $G(i) = (V(i), E(i))$, we obtain a subgraph $G' = (V', E')$ in which all pairs of the same vertices are joined and renamed as $\tau_l$, for $l = 1,2,...$ Set $V'$ differs from $V(i)$ in having $\tau_l$ vertices instead of the pairs of the vertices with the same name (originally belonging to silent moves in the framework). Hence, we can conclude that $K_B = V' \setminus (\{i\} \cup \tau_l)$. We shall call the algorithm explained above 'Removing Keys from a Set', denoted as Rem. We shall write $\text{Rem}(K_F) = K_B$, meaning that $K_B$ can be obtained by applying algorithm Rem to $K_F$.

Before stating and proving a lemma about structural correspondence, we need to introduce an erasing function $\lambda$ (the function $\varphi$ is defined in Section 4.1):

**Definition 28** (Erasing function for causal processes)**.** *The erasing function $\lambda$ that maps causal processes from Boreale and Sangiorgi's semantics to $\pi$-calculus is inductively defined as follows:*

$$\lambda(A \mid A') = \lambda(A) \mid \lambda(A') \qquad \lambda(K_B :: A) = \lambda(A)$$
$$\lambda(\nu a(A)) = \nu a(\lambda(A)) \qquad \lambda(P) = P$$

Now we can show our structural correspondence.

**Lemma 10** (Structural correspondence)**.** *Starting from initial $\pi$-calculus process $P$, where $P = A_1 = X_1$, we have:*

1. *if $P \xrightarrow{\zeta_1}_{K_{B1}} A_2 \ldots A_n \xrightarrow{\zeta_n}_{K_{Bn}} A_{n+1}$ then there exists a trace $P \xrightarrow{\mu_1}_{K_{F1}} X_2 \ldots X_n \xrightarrow{\mu_n}_{K_{Fn}} X_{n+1}$ and $K_{Fi}$, such that for all $i$, $\lambda(A_i) = \varphi(X_i)$, $\zeta_i = \gamma(\mu_i)$ and $\text{Rem}(K_{Fi}) = K_{Bi}$, for $i = 1,...,n$.*

2. if $P \xrightarrow[K_{F1}]{\mu_1} X_2 \ldots X_n \xrightarrow[K_{Fn}]{\mu_n} X_{n+1}$ then there exists a trace $P \xrightarrow[K_{B1}]{\zeta_1} A_2 \ldots A_n \xrightarrow[K_{Bn}]{\zeta_n} A_{n+1}$ where for all $i$, $\lambda(A_i) = \varphi(X_i)$, $\zeta_i = \gamma(\mu_i)$ and $\text{Rem}(K_{Fi}) = K_{Bi}$, for $i = 1, \ldots, n$.

*Proof.* Both directions (*1.* and *2.*) are proved by induction on the length of the computation. Let us consider direction *1*.

(I) The base case is given by a single transition, and there is no cause; hence, $K_{B1} = K_{F1} = \emptyset$. We proceed by induction on the structure of the π-calculus process $P$ and the last applied rule on the transitions $t$ and $t'$, where $t : P \xrightarrow{\zeta_1} A_2$ and $t' : P \xrightarrow{\mu_1} X_2$.

- $P = \pi.P'$ where $\pi = \overline{b}a$ or $\pi = b(x)$; Rules that can be applied in Boreale and Sangiorgi's semantics are BS-OUT and BS-IN. We shall show the case when rule BS-OUT is applied; the other case is similar. We have $\pi.P' \xrightarrow{i_1 : \pi} \{i_1\} :: P' = A_2$ where $\lambda(A_2) = P'$. In the framework we can execute the corresponding action by applying the rule OUT1 and we have $\pi^*.P' \xrightarrow{(i_1,*,*):\pi} \pi^*[i_1,*].P' = X_2$ with $\varphi(X_2) = P'$ as desired.

- $P = Q \mid Q'$; Rules that can be applied in Boreale and Sangiorgi's semantics are BS-PAR, BS-COM and BS-CLOSE. We shall show the case when rule BS-CLOSE is used; the rest of the cases are similar. Since rule BS-CLOSE is applied on the process $Q \mid Q'$ one of the parallel components needs to extrude a bound name. Let it be a process $Q = va(Q_1)$. Then we have

$$va(Q_1) \mid Q' \xrightarrow{\tau} va(Q_1' \mid Q''\{a/x\}) = A_2$$

with the premises $va(Q_1) \xrightarrow{i_1 : \overline{b}\langle va \rangle} Q_1'$ and $Q' \xrightarrow{i_1 : b(x)} Q''$, for some $b$. Since there are no causes, we have $\lambda(A_2) = va(Q_1' \mid Q''\{a/x\})$.

In the framework we can execute the corresponding synchronisation by applying the rule CLOSE and we have:

$$va_{\emptyset_*}(Q_1) \mid Q' \xrightarrow{(i_1,*,*):\tau} va_{\emptyset_*}(va_{\{i_1\}_*}(Y_1') \mid Y''\{a^{i_1}/x\}) = X_2$$

with the premises $va_{\emptyset_*}(Q_1) \xrightarrow{(i_1,*,*):\overline{b}\langle va \rangle} va_{\{i_1\}_*}(Y_1')$, where $\varphi(Y_1') = Q_1'$; and $Q' \xrightarrow{(i_1,*,*):b(x)} Y''$ where $\varphi(Y'') = Q''$. Then we have

$$\varphi(va_{\emptyset_*}(va_{\{i_1\}_*}(Y_1') \mid Y''\{a^{i_1}/x\})) = va(Q_1' \mid Q''\{a/x\})$$

as desired.

- $P = va(P')$; Rules that can be applied in Boreale and Sangiorgi's semantics are BS-RES and BS-OPEN. We shall show the case when rule BS-OPEN is applied; the other case is similar to the one above. If the rule BS-OPEN is applied on the process $va(P')$, executed action extrudes name $a$, and we have:

$$va(P') \xrightarrow{i_1 : \overline{b}\langle va \rangle} \{i_1\} :: P'' = A_2$$

with the premise $P' \xrightarrow{i_1 : \overline{b}a} \{i_1\} :: P''$. By discarding cause set $\{i_1\}$ we have $\lambda(A_2) = P''$.

We can match the same action in the framework and apply rule OPEN on the process $va_{\emptyset_*}(P')$ and obtain:

$$va_{\emptyset_*}(P') \xrightarrow{(i_1,*,*):\overline{b}\langle va \rangle} va_{\{i_1\}_{i_1}}(Y') = X_2$$

with the premise $P' \xrightarrow{(i_1,*,*):\overline{b}a} Y'$ where $\varphi(Y') = P''$. By discarding the elements of the history from the process $X_2$, we have $\varphi(X_2) = P''$ as desired.

(II) In the inductive case we let $s_{BS}: A_1 \xrightarrow[K_{B1}]{\zeta_1} A_2 \ldots A_n \xrightarrow[K_{Bn}]{\zeta_n} A_{n+1}$ be the trace on causal processes and $s_F: X_1 \xrightarrow[K_{F1}]{\mu_1} X_2 \ldots X_n \xrightarrow[K_{Fn}]{\mu_n} X_{n+1}$ the trace in the framework, where $A_1 = X_1 = P$; and let us suppose that the inductive hypothesis holds for these two traces. By inductive hypothesis, we have that $\lambda(A_i) = \varphi(X_i) = P_i$ and in the framework there exist sets $K_{Fi}$, such that $\text{Rem}(K_{Fi}) = K_{Bi}$ for all $i = 1, \ldots, n$.

To prove the inductive step, let

$$t: s_{BS} \xrightarrow[K_{Bn+1}]{\zeta_{n+1}} A_{n+2} \quad \text{and} \quad t': s_F \xrightarrow[K_{Fn+1}]{\mu_{n+1}} X_{n+2}$$

be two corresponding computations. We need to prove two statements: (1) $\text{Rem}(K_{Fn+1}) = K_{Bn+1}$ and (2) $\lambda(A_{n+2}) = \varphi(X_{n+2})$.

To prove (1) we should look at the action $\zeta_n$ because it is the last action that can influence the cause set $K_{Bn+1}$ (cause set $K_{Bn+1}$ does not depend on the action $\zeta_{n+1}$). There are two main cases:

- action $\zeta_n$ is the direct structural cause of the action $\zeta_{n+1}$; we have that action $\zeta_n$ is a visible action and $K_{Bn+1} = K_{Bn} \cup \{i_n\}$. By inductive hypothesis, we have that there exist $K_{Fn}, \mu_n \in s_F$ such that $\gamma(\mu_n) = \zeta_n$ and $\text{Rem}(K_{Fn}) = K_{Bn}$. The action $\mu_n$ is identified with a key $i_n$ and it is a visible one; therefore we have $K_{Fn+1} = K_{Fn} \cup \{i_n\}$. The $\text{Rem}$ algorithm does not remove keys of the visible actions and we have $\text{Rem}(K_{Fn} \cup \{i_n\}) = K_{Bn} \cup \{i_n\} = K_{Bn+1}$ as desired.

- action $\zeta_n$ is not the direct cause of the action $\zeta_{n+1}$; then $\zeta_n = \tau$ or $\zeta_n$ happened on a different component in the parallel composition from the action $\zeta_{n+1}$.

  - If $\zeta_n = \tau$, there exist $K_{Bj}, K_{Bh} \in s_{BS}$ such that $K_{Bj}$ is a cause set of the input action and $K_{Bh}$ is a cause set of the output action which communicate in a $\tau$ move. Since $\tau$ actions merge cause sets we have that $K_{Bn+1} = K_{Bj} \cup K_{Bh}$. In the framework, a $\tau$ move is composed of the same input and output actions as in the trace on the causal processes. Hence, there exist $K_{Fj}, K_{Fh} \in s_F$, and by inductive hypothesis $\text{Rem}(K_{Fj}) = K_{Bj}$ and $\text{Rem}(K_{Fh}) = K_{Bh}$. Since in the framework a $\tau$ action is identified with a key $i_n$ we have that $K_{Fn+1} = K_{Fj} \cup K_{Fh} \cup \{i_n\}$. By definition, the $\text{Rem}$ algorithm removes keys belonging to the $\tau$ actions, and we have $\text{Rem}(K_{Fj} \cup K_{Fh} \cup \{i_n\}) = K_{Bn+1}$ as desired.

  - If $\zeta_n$ happened on a different component in the parallel composition, there exist $K_{Bh+1}, \zeta_h \in s_{BS}$ where $\zeta_h$ is the last action on the same component in the parallel composition as $\zeta_{n+1}$ ($K_{Bh} :: P \mid A \xrightarrow[K_{Bh}]{\zeta_h} K_{Bh+1} :: P' \mid A$). Then we have that $K_{Bn+1} = K_{Bh+1}$, since $\zeta_h$ was the last action before $\zeta_{n+1}$. By inductive hypothesis, in the framework, there exist $K_{Fh+1}, \mu_h \in s_F$, where $\gamma(\mu_h) = \zeta_h$ and $\text{Rem}(K_{Fh+1}) = K_{Bh+1}$. By the same observation we have $K_{Fh+1} = K_{Fn+1}$ as desired.

We prove case (2) by induction on the structure of the $\pi$-calculus process $P$, where $\lambda(A_{n+1}) = \varphi(X_{n+1}) = P$ and the last applied rule on the transitions $t$ and $t'$. The reasoning is similar to that for the base case. (Using the fact that $\text{Rem}(K_{Fn+1}) = K_{Bn+1}$, we ensure that for an action $\zeta_{n+1}$ there exists just one corresponding action $\mu_{n+1}$, such that $\gamma(\mu_{n+1}) = \zeta_{n+1}$.)

**Remark 3.** The rule BS-CAU inductively allows a causal process $A = K_B :: P$ to execute if process $P$ can execute, while the executed action brings its cause set $K_B$. In the framework, it is done with the rules IN2 and OUT2, which inductively allow a reversible process to move, independent of its history.

Direction *2.* is proved by induction on the length of the computation (similarly to *1.*).

(I) The base case is given by a single transition, and we have that there is no cause, hence, $K_{F1} = K_{B1} = \emptyset$. We proceed by induction on the structure of the $\pi$-calculus process $P$ and the last applied rule on the transitions $t$ and $t'$, where $t : P \xrightarrow{\mu_1} X_2$ and $t' : P \xrightarrow{\zeta_1} A_2$. We show correspondence between the rules that can be applied on process $P$, similarly to *1*.

(II) In inductive case we let $s_{BS}$ and $s_F$ to be defined as in the case above with $A_1 = X_1 = P$. We suppose that the inductive hypothesis holds for these two traces and we have $\varphi(X_i) = \lambda(A_i) = P_i$ and $\text{Rem}(K_{Fi}) = K_{Bi}$ for all $i = 1, \ldots, n$.

To prove the inductive step, let $t : s_F \xrightarrow[K_{Fn+1}]{\mu_{n+1}} X_{n+2}$ and $t' : s_{BS} \xrightarrow[K_{Bn+1}]{\zeta_{n+1}} A_{n+2}$ be two corresponding computations. We need to prove two statements: (1) $\text{Rem}(K_{Fn+1}) = K_{Bn+1}$ and (2) $\varphi(X_{n+2}) = \lambda(A_{n+2})$.

To prove (1) we should look at the action $\mu_n$ because it can influence cause set $K_{Fn+1}$ (cause set $K_{Fn+1}$ does not depend on the action $\mu_{n+1}$). There are three cases:

- action $\mu_n$ is the direct structural cause of the action $\mu_{n+1}$ and it is a visible action; then we have $K_{Fn+1} = K_{Fn} \cup \{i_n\}$. By inductive hypothesis, we have that there exist $K_{Bn}, \zeta_n \in s_{BS}$ such that $\gamma(\mu_n) = \zeta_n$ and $\text{Rem}(K_{Fn}) = K_{Bn}$. Since $\mu_n$ is visible, $\zeta_n$ needs to be too and it is identified with the key $i_n$. We have $K_{Bn+1} = K_{Bn} \cup \{i_n\}$ as desired.

- action $\mu_n$ is the direct structural cause of the action $\mu_{n+1}$ and it is a silent action. In the framework we have $K_{Fn+1} = K_{Fn} \cup \{i_n\}$, since silent action is identified with the key $i_n$. Cause set $K_{Fn}$ of the $\tau$ action contains cause sets of the communicating actions (input and the output ones). Hence, there exist $K_{Fj}, K_{Fh} \in s_F$ such that $K_{Fn} = K_{Fj} \cup K_{Fh}$. By inductive hypothesis, we know that there exist $K_{Bj}, K_{Bh} \in s_{BS}$ such that $\text{Rem}(K_{Fj}) = K_{Bj}$ and $\text{Rem}(K_{Fh}) = K_{Bh}$. Since silent actions on causal processes just merge two cause sets, we have $K_{Bn+1} = K_{Bj} \cup K_{Bh}$. In the algorithm Rem, keys belonging to $\tau$ actions are removed form the cause set in the framework, and we have $K_{Bn+1} = \text{Rem}(K_{Fn} \cup \{i_n\}) = \text{Rem}(K_{Fn}) = K_{Bj} \cup K_{Bh}$ as desired.

- action $\mu_n$ is not the direct cause of the action $\mu_{n+1}$; then $\mu_n$ happened on a different component in the parallel composition from the action $\mu_{n+1}$. In this case, there exist $K_{Fh+1}, \mu_h \in s_F$ where $\mu_h$ is the last action on the same component in the parallel composition as $\mu_{n+1}$. Hence, action $\mu_h$ is direct cause of the action $\mu_{n+1}$ and we have the same reasoning as in the cases above.

We prove case (2) by induction on the structure of the π-calculus process $P$, where $\varphi(X_{n+1}) = \lambda(A_{n+1}) = P$ and last applied rule on the transitions $t$ and $t'$. The reasoning is similar to the base case. (Using the fact that $\text{Rem}(K_{Fn+1}) = K_{Bn+1}$, we ensure that for an action $\mu_{n+1}$ there exists just one corresponding action $\zeta_{n+1}$, such that $\gamma(\mu_{n+1}) = \zeta_{n+1}$.)  □

**Remark 4.** In the framework, rule CAUSE REF can be applied to update cause set $K$ of the action using an extruded name in the subject position. This rule does not influence the structure of the process $X$; it just records actions that have extruded bound names.

The object causality in Boreale and Sangiorgi's semantics is defined on the trace of a process. The first action that extrudes a bound name will cause all the future actions using that name in any position of the label. We adapt the definition of object causality to the late semantics where we use traces defined on the causal process $A$.

**Definition 29** (object causality). *In a trace $A_1 \xrightarrow[K_{B1}]{\zeta_1} A_2 \cdots A_n \xrightarrow[K_{Bn}]{\zeta_n} A_{n+1}$ where $A_1$ is a π-calculus process $P$, if*

- $\zeta_i = \overline{b}\langle va \rangle$ *where $a \cap \text{fn}(A_i) = \emptyset$ and for all $j < i$, $a \cap \text{n}(\zeta_j) = \emptyset$ we say that name b is* introduced *in $\zeta_i$. Action $\zeta_h$ is* object dependent *on $\zeta_i$, $1 \leqslant i < h \leqslant n$, if there is a name introduced in $\zeta_i$ which is among the free names of $\zeta_h$.*

- $\zeta_i = b(x)$ where $x \cap \mathtt{fn}(A_i) = \emptyset$ *and for all* $j < i$, $x \cap \mathtt{n}(\zeta_j) = \emptyset$ *we say that variable x is* introduced in $\zeta_i$. *Action* $\zeta_h$ *is* object dependent *on* $\zeta_i$, $1 \leqslant i < h \leqslant n$, *if there is a variable introduced in* $\zeta_i$ *which is among the free variables of* $\zeta_h$.

We can define object causality induced by the framework, on the trace of a reversible process. In this way Definition 18 will be the case when $n = 2$.

**Definition 30** (Object causality on the trace). *In the trace* $t_1 : X_1 \xrightarrow{(i_1,K_1,j_1):\pi_1} X_2 \cdots t_n : X_n \xrightarrow{(i_n,K_n,j_1):\pi_n} X_{n+1}$, *transition* $t_h$ *is an* object cause *of transition* $t_l$, *written* $t_h < t_l$, *if:*

- $i_h \in K_l$, *when we consider two forward transitions* $t_h$ *and* $t_l$ *or*
- $i_l \in K_h$, *when we consider two backward transitions* $t_h$ *and* $t_l$

*We assume that* $t_h \neq t_l^\bullet$.

The next theorem will prove causal correspondence between causality in the framework when memory $\Delta$ is instantiated with $\Gamma_w$ and Boreale and Sangiorgi's late causal semantics.

**Causal correspondence (Theorem 2).** *The reflexive and transitive closure of causality introduced in [6] coincides with the causality of the framework when* $\Delta = \Gamma_w$.

*Proof.* The proof relies on Lemma 10 and the fact that object dependence induced by input action in Boreale and Sangiorgi's semantics is subject dependence as well. By design of the framework and definitions for predicates $\mathtt{Cause}(\cdot)$ and $\mathtt{Update}(\cdot)$ the first extrusion of a name will cause every other action using that name (this is accomplished with the rules OPEN and CAUSE REF). In Definition 29 object dependence induced by an input action is also the structural one, and the one induced by extrusion coincides with object dependence in the framework. □